\documentclass[10pt,twocolumn]{article}

\usepackage{graphicx}
\usepackage{amsfonts}
\usepackage{amsmath}
\usepackage{amssymb}
\usepackage{mathrsfs} 
\usepackage{lipsum}
\usepackage{color}
\usepackage[squaren,Gray,cdot]{SIunits}
\usepackage{authblk}
\usepackage[top=2.5cm, bottom = 2.5cm, left = 2cm, right = 2cm]{geometry}
\usepackage{mathtools, cuted}
\usepackage{array,amsmath,booktabs}
\usepackage[font=small, labelfont={bf,normalsize}]{caption}
\usepackage{subcaption}
\usepackage[hyphens]{url}
\usepackage{multirow}
\usepackage{hyperref}
\usepackage{rotating}

\usepackage{sectsty}
\sectionfont{\sffamily}
\subsectionfont{\sffamily}
\subsubsectionfont{\sffamily}
\paragraphfont{\sffamily}

\newcommand{\cSt}{\centi \mathrm{St}}
\newcommand{\N}{\mathcal{N}}
\newcommand{\Aprime}{\mathcal{A}^{\prime}}
\newcommand{\Dh}{\mathcal{D}_{\mathrm{eq}}}
\newcommand{\Mone}{\langle d \rangle}
\newcommand{\Mtwo}{\langle d^2 \rangle}
\newcommand{\Dwet}{D_{B,\mathrm{wet}}}
\newcommand{\Ddry}{D_{B,\mathrm{dry}}}

\title{\sf The Fluid Mechanics of Splatter Painting}

\author[1]{Diego \'{A}vila-Garc\'{i}a}
\author[1]{Carlos Salazar-Mart\'{i}n}
\author[1,2]{Javier Rodr\'{\i}guez-Rodr\'{\i}guez}
\author[3,4]{Mithun Ravisankar}
\author[3]{Roberto Zenit}
\author[1,2,5]{Lorène Champougny}
\affil[1]{Universidad Carlos III de Madrid, Department of Thermal and Fluids Engineering, Avenida de la Universidad, 30 (edificio Sabatini), 28911 Leganés (Madrid), Spain}
\affil[2]{Universidad Carlos III de Madrid, Instituto Gregorio Mill\'an, Avenida de la Universidad, 30 (edificio Sabatini), 28911 Leganés (Madrid), Spain}
\affil[3]{Center for Fluid Mechanics, Brown University, Providence, RI 02912, USA}
\affil[4]{Department of Mathematics, Mechanics division, University of Oslo, Oslo 0316, Norway}
\affil[5]{University of Toulouse, Toulouse INP, CNRS, LGC, Toulouse, France}
\date{}
\begin{document}

%
\twocolumn[
\begin{@twocolumnfalse}
    \maketitle

    \begin{abstract}
        In splatter painting, a collection of liquid droplets is projected onto the substrate by imposing a controlled acceleration to a paint-loaded brush.
        This work aims at unraveling the physical phenomena at play in this widespread artistic technique.
        By characterizing the kinematics of the splattering action (brush flicking or tapping), we identify the main force driving liquid out of the tip of the brush (hair bundle).
        For real or simplified brushes, the amount of liquid expelled is measured for various liquid viscosities and imposed accelerations, among other parameters.
        Experimental trends are successfully captured by a physical model that reproduces two distinguished limits: an inertia-dominated flow and a viscous-dominated flow in an anisotropic porous medium with parallel pores.
        Beyond splatter painting, our analysis provides a framework to describe mechanical drying strategies for fibrous materials or porous structures.
    \end{abstract}
    \vspace{0.5cm}
\end{@twocolumnfalse}
]
%
%
Painting is one of the earliest human artistic expressions.
Early cave paintings are considered by many the signature for the emergence of human behavior \cite{GROSS2020}. 
A series of recent papers have argued that the essence of artistic painting lies within a keen understanding of the physical constraints imposed by the interaction between the artist, their instruments and materials \cite{Herczynski2011,Zetina2015,Palacios2019,Zenit2019,Sun2024}. 
In other words, artists develop a deep empirical knowledge of the physical mechanisms that allow them to manipulate fluid flows to paint. 
They skillfully choose tools, paints and action to cover surfaces with certain fluid textures in order to convey their intention to the painting.

\begin{figure}[t]
    \centering
    \includegraphics[width = 0.9\linewidth]{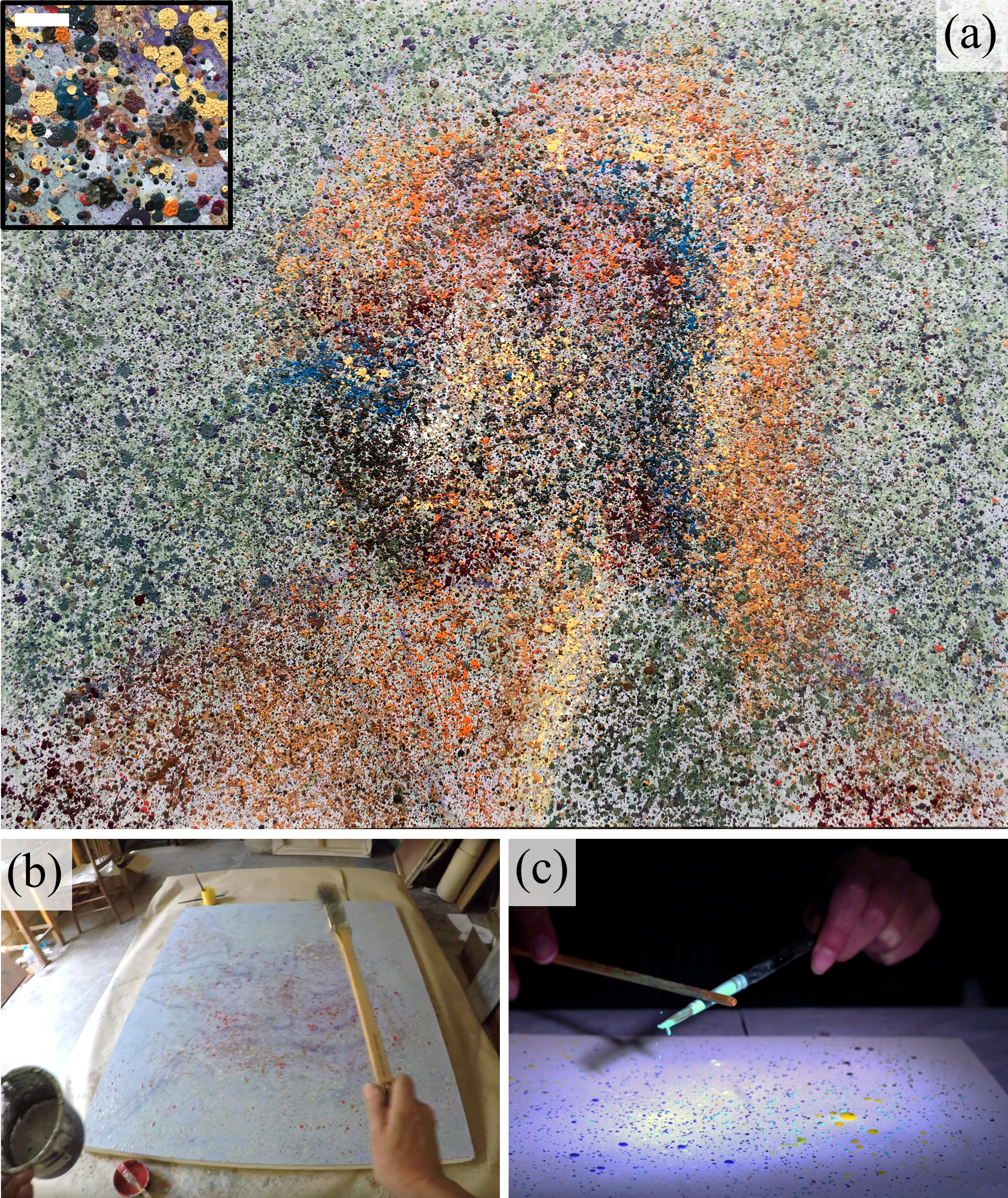}
\caption{\textbf{Splatter painting technique.} (a) Painting produced by splatter painting: \textit{Portrait of Nadia Cortes} by Octavio Moctezuma, 2022 (reproduced with permission). Inset: Zoom of the droplet pattern (Scale bar: $1~\centi\meter$) --- (b) Snapshot of a point-of-view movie of artist O. Moctezuma splattering paint by flicking the brush (Movie S1). --- (c) Snapshot of artist C. Champougny splattering paint by tapping the brush (Movie S4).}
\label{fig:artwork_actions}
\end{figure}

In this work we investigate the splatter painting technique. 
It consists in subjecting a paint-loaded brush to a sudden acceleration, for example by flicking the brush (Figure \ref{fig:artwork_actions}b) or tapping its handle (Figure \ref{fig:artwork_actions}c).
As a result, paint droplets or filaments are expelled from the brush and projected onto a substrate -- often laid horizontally -- where they leave patterns of drops or lines \cite{Herczynski2011}.
Figure \ref{fig:artwork_actions}a displays an example of a modern artwork created with splatter painting; Movie S1 shows the process in action.
In this technique, perhaps most emblematically used by American painters Jackson Pollock (1912-1956) \cite{Zenit2019} and Sam Francis (1923-1994) \cite{Burchett2019}, the brush never touches the canvas.
Splatter painting is thus considered an \emph{action painting} technique. 
This term, coined by Harold Rosenberg \cite{rosenberg1952}, refers to the use of dribbling, splashing or smearing paint onto a canvas. 
Its relevance, discussed by many art historians and critics, emerges from the fact that the creation process is not entirely under the control of the artist; the deposition of paint on the canvas is governed by the mechanical constraints of the process as much as it is by the will of the artist. 
In this sense, Rosenberg argued that art could be redefined to include the act rather than just the object, as a process rather than a product.
In splatter painting specifically, the artist selects a set of conditions: paint properties (mostly viscosity), action (the intensity of flicking or tapping) and tools (type and size of brush).
Paint deposition is then the result of the spontaneous evolution of the system, given this set of conditions.

The splatter painting technique can be conceptualized as being the sequence of the following physical processes.
(i) First, the brush is dipped in paint: a certain volume of liquid is retained inside the bundle of bristles without dripping, due to (visco-) capillary effects \cite{Princen1969a, Princen1969b, Charpentier2020, Radisson2025}. 
(ii) Once loaded, the brush is accelerated, either by rapid hand motion or a handle tapping, causing the fluid trapped within the bristles to flow and eventually leave the brush.  
Acceleration-driven ejection of the liquid from pores is therefore the key element of this technique.
In nature, generating large accelerations by shaking is a common strategy to remove undesired moisture from small gaps: wet mammals do it to dry their fur \cite{Dickerson2012}, humans to get rid of water in the ear canal \cite{Kim2023} for instance.
In industrial processes, centrifugal forces are commonly used to dry porous materials, such as fabrics \cite{anlauf2007} or 3D-printed structures \cite{cheng2022}.
While the uptake of fluids in porous or hairy structures is the object of active research \cite{gambaryan2014, Charpentier2020, Cheng2024, Siefert2025, Radisson2025}, many aspects of the acceleration-driven drying process remain to be addressed -- some discussed here. 
(iii) Once the fluid is ejected from the bristles, it forms filaments that fly away. 
In most cases, these filaments fragment into droplets due to the classical Rayleigh-Plateau instability \cite{Villermaux2020}, more prominently in low viscosity liquids. 
This strategy is used in industrial applications to atomize liquids \cite{Keshavarz2020}.
(iv) After the droplets are formed, they land on the canvas, usually laid horizontal, to create the artist composition, as illustrated in Fig. \ref{fig:artwork_actions}. 
This final step -- droplet impact on a substrate -- has been extensively studied \cite{yarin2006, josserand2016}, due to its importance in many engineering applications and natural phenomena, from inkjet printing \cite{Lohse2022} to pathogens transmission \cite{Bourouiba2021ARFM}. 

The process of splatter painting is clearly a rich fluid mechanical problem. 
In this work, we focus our attention on how and how much liquid is ejected from the bristles upon sudden acceleration.
It is both the most critical step in splatter painting, and the least understood in literature.
To unravel the physical mechanisms at play, we reproduce the tapping action with a dedicated setup allowing us to impose a controlled acceleration to a liquid-loaded brush or tube. 
We compose a model that captures the essential features of our experimental data, and show that our results can be extended to more complex splatter painting actions, such as flicking.
%
\section{Kinematics of splatter painting actions} \label{sec:kinematics}
%
\subsection{Brush flicking} \label{ssec:flicking}
%
\begin{figure}[t]
    \centering
    \includegraphics[width = \linewidth]{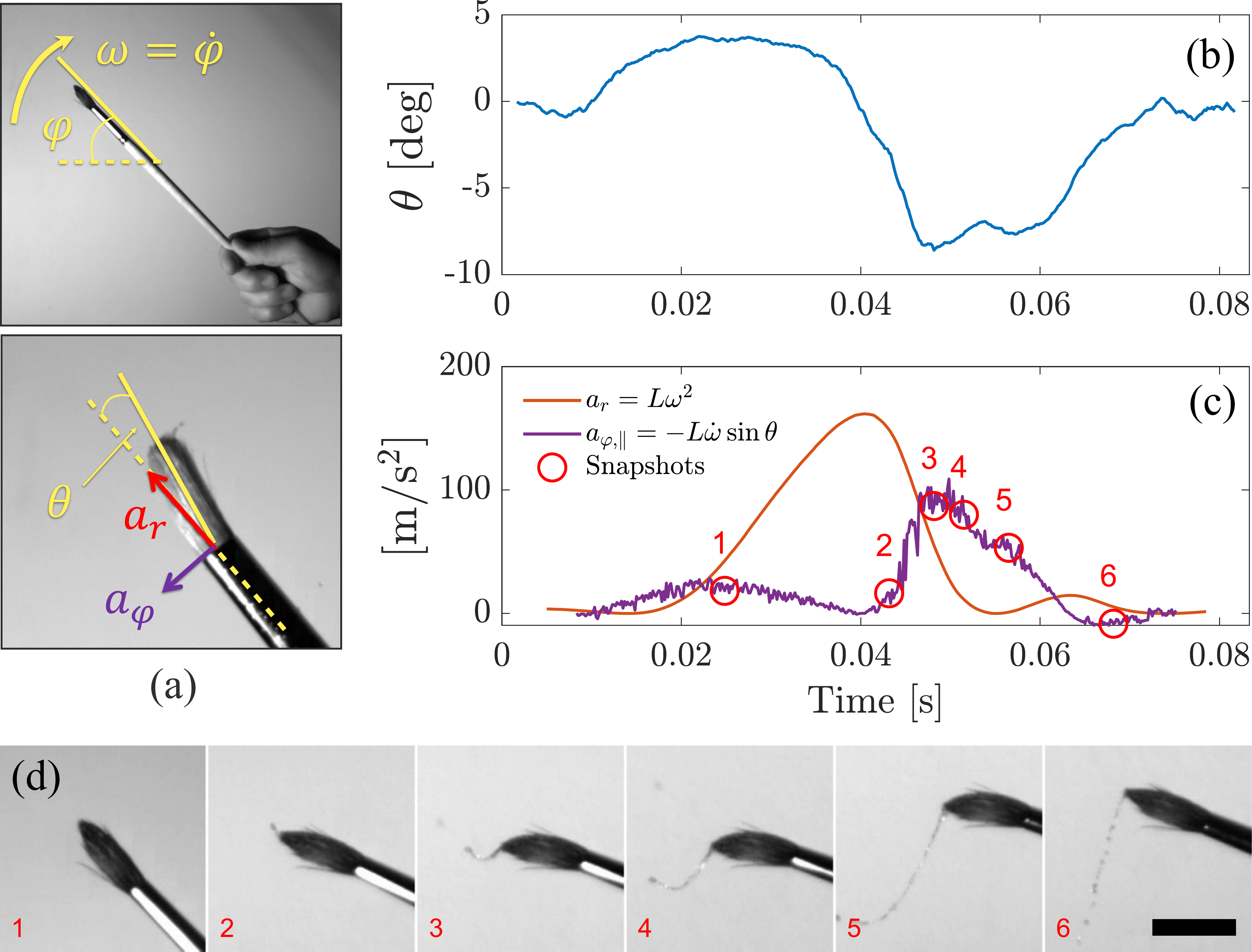}
    \caption{\textbf{Kinematics of brush flicking.} (a) Image of the brush being flicked, with the definition of the angle $\varphi$ and the angular speed $\omega$. The inset displays another brush tip to illustrate the definition of the angle $\theta$. (b) Evolution of $\theta$ for a typical flicking event. The time axis is synchronized with panel (c). (c) Centrifugal and tangential accelerations at the brush's tip (Brush length: 30 cm). Red circles correspond to the snapshots of the brush's tip shown in panel (d). Scale bar: 20 mm. See Movie S2.}
\label{fig:kinematics_flicking}
\end{figure}
We start by investigating the kinematics of what is perhaps the most natural splatter painting action: brush flicking.
It consists in imparting angular momentum to the brush for a brief moment, using a whipping motion of the wrist.
As illustrated in Fig. \ref{fig:kinematics_flicking}a, and in Movie S2, the motion of the brush can be primarily described by the angle $\varphi$ of the handle with respect to a fixed direction (say horizontal).
The angular speed is then simply $\omega = \dot{\varphi}$, and we denote $L$ the total length of the brush (handle plus bundle).

In a reference frame that moves with the bundle, the total acceleration applied to the paint has two components: the centrifugal acceleration $a_r = L\omega^2$ in the radial direction, and the acceleration $a_{\varphi} = -L\dot{\omega}$ due to the time variation of $\omega$ in the orthoradial direction. 
However, when it comes to displacing liquid inside an anisotropic medium such as a bundle of fibers, these accelerations are not equivalent. 
From the bundle geometry, we can estimate that the transversal permeability (\textit{i.e.} perpendicular to the bristles) is about 5-10 times smaller than the longitudinal one (see Section \ref{apdx:properties_porous} in SI).
It is therefore easier for paint to move in the longitudinal direction, which is why we now restrict our discussion to the acceleration components along that direction.

If the bristles were rigid, the longitudinal direction of the bundle would coincide with the radial direction.
For a flexible bundle, the orthoradial acceleration $a_{\varphi}$ slightly bends the bristles, causing a misalignment $\theta$ between the handle and the bundle (Fig. \ref{fig:kinematics_flicking}a).
As a consequence, the orthoradial acceleration has a finite projection   $a_{\varphi, \parallel} = - L\dot{\omega}\,\sin\theta$ along the bundle's axis.
The angle $\theta$ remains small ($<10\deg$, Fig. \ref{fig:kinematics_flicking}b), so the centrifugal acceleration nearly coincides with its projection along the bundle: $a_{r,\parallel} = a_r \cos \theta \approx a_r$. 

In Fig. \ref{fig:kinematics_flicking}c we display the evolution of the two acceleration contributions that project onto the bundle's axis: $a_r = L \omega^2$ and $a_{\varphi,\parallel} = - L\dot{\omega}\,\sin\theta$. 
The snapshots in panel (d) correspond to the times marked with circles in (c). 
We observe that liquid ejection starts close to the time when $a_r$ is maximum and continues until both accelerations have decayed to a value much lower than their peaks. 
In fact, the liquid ligament shown in snapshot 6 of Fig.~\ref{fig:kinematics_flicking}d is the same one as in snapshot 5 but atomized into droplets (Movie S2), which means that ejection has already stopped.
It is worth noting that, although the orthoradial contribution $a_{\varphi,\parallel}$ is smaller and lasts for a shorter time than the centrifugal acceleration, it seems to also contribute to liquid ejection. This may be due to the fact that $a_{\varphi}$ not only causes the deflection of the bristles, but also opens up the bundle, as can be observed by comparing, for instance, snapshots 3 (maximum $a_{\varphi}$) and 6, when $\dot{\omega}$ is much smaller. 
This widens the space between the bristles, thus allowing liquid to flow more easily.

From this experiment, we conclude that that liquid ejection coincides with the moment of maximum acceleration along the bundle direction. 
Clearly, both the centrifugal acceleration $a_r$ and the longitudinal projection of the orthonormal acceleration $a_{\varphi,\parallel}$ contribute -- but it is unclear in which proportions.
Hence, we now turn to a different splatter painting action -- namely brush tapping -- for which only one of these contributions is present, therefore allowing for a more precise assessment of the mechanism of paint ejection. 
%
\subsection{Brush tapping} \label{ssec:tapping}
%
\begin{figure*}[t]
    \centering
    \includegraphics[width = \linewidth]{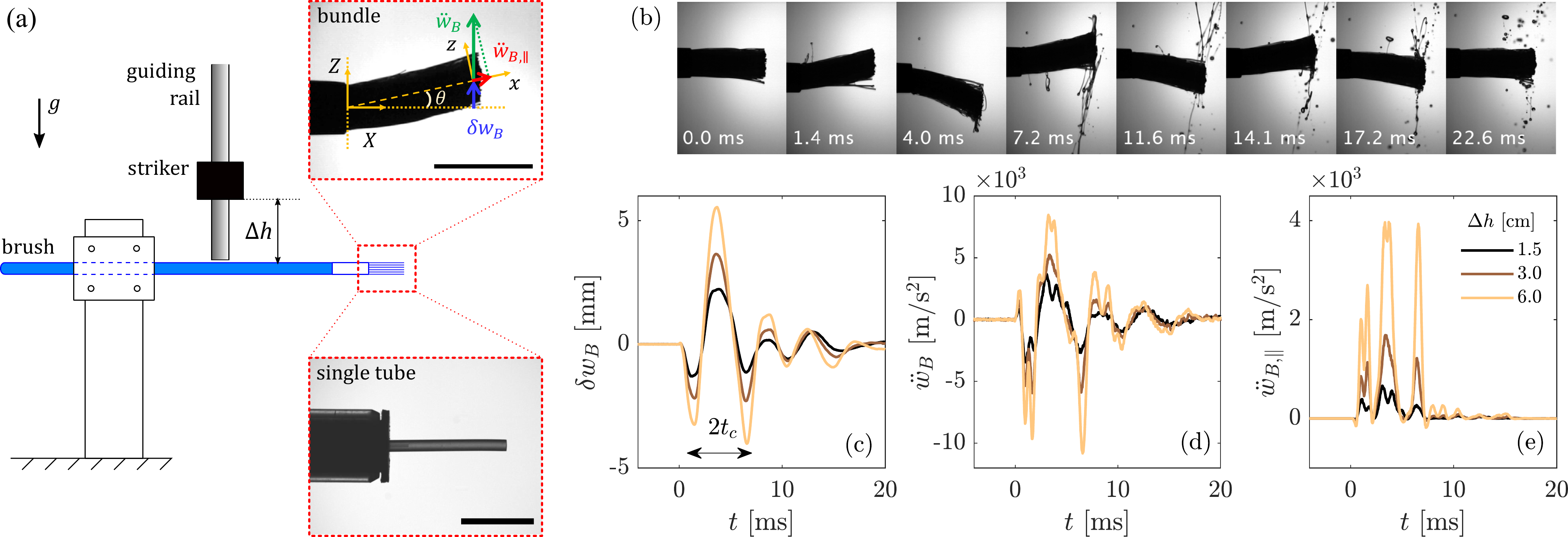}
    \caption{(a) Experimental setup emulating the tapping action used in splatter painting. The brush handle is hit by a striker dropped from a height $\Delta h$. As a result, the flexible tip of the brush (hair bundle in the case of a real brush, or tube in the case of the single-pore simplified system) is set in motion. Scale bars on pictures: $1~\centi\meter$. --- (b) Snapshot sequence of the oscillating bundle upon impact (liquid SO-40, drop height $\Delta h = 3~\centi\meter$); se also Movie S3. --- Image analysis allows us to track as a function of time (c) the bundle deflection $\delta w_B$, (d) the vertical acceleration $\ddot{w}_{B}$ in the reference frame of the laboratory, and (e) its projection $\ddot{w}_{B,\parallel}$ onto the longitudinal axis of the bundle. These data correspond to a position $X_{\mathrm{tip}} = 0.85 L_B$ along the bundle of a brush loaded with $1000~\cSt$ silicone oil.}
    \label{fig:setup_kinematics}
\end{figure*}
Experimentally, we emulate the brush tapping action by dropping a weight (striker, mass $M$) from a given height $\Delta h$, onto the brush handle (Fig.~\ref{fig:setup_kinematics}a, see and Materials and Methods for details).
How the impact of a mass falling from a given height translates into the brush motion is a classical problem in impact mechanics, which is not fully address here. 
The interested reader is referred to the Supporting Information (SI, Section \ref{apdx:impact_mechanics}) for more details about this problem and how it can be modeled.
Instead, we focus directly on the kinematics of the brush after impact, as revealed by high-speed imaging (see Fig.~\ref{fig:setup_kinematics}b and supplementary Movie S3).
In order to quantify the movement of the brush, we introduce the $(X,Z)$ coordinate system tied to the (non-inertial) reference frame of the accelerated brush handle.
Digital image processing of sequences, such as that shown in Fig.~\ref{fig:setup_kinematics}b, allows us to measure vertical displacements in the reference frame of the laboratory: $w_H(t)$ for the rigid handle, and $w_B (X,t)$ for the deformable bundle.
The bundle vertical displacement relative to the handle, that is to say its deflection with respect to its equilibrium position, is defined as $\delta w_B(X,t) = w_B(X,t) - w_H(t)$ and plotted in Fig.~\ref{fig:setup_kinematics}c.
It shows a few damped oscillation cycles, with typical period $2t_c$, before the bundle goes back to its initial position.
The vertical displacement $w_B$ is time-differenciated twice to obtain the instantaneous vertical acceleration $\ddot{w}_B(X,t)$ of the bundle (Fig.~\ref{fig:setup_kinematics}d) in the reference frame of the laboratory.

For our experimental conditions, we observe that liquid is predominantly ejected from the tip of the bundle, as shown in Fig.~\ref{fig:setup_kinematics}b.
Assuming an initially uniform liquid distribution in the bristles, this observation suggests that the vertical acceleration $\ddot{w}_B$ mostly drives liquid \emph{along} the bundle to the tip, where it detaches and fragments.
At each point along the central line of the bundle, the local, vertical acceleration $\ddot{w}_B$ can be divided into a transversal and a longitudinal component (Fig.~\ref{fig:setup_kinematics}a).
Because of the top-down symmetry of the motion, the former averages out to zero over several oscillation cycles, while the latter, denoted $\ddot{w}_{B,\parallel}$, imposes an acceleration with non-zero average to the fluid.
Observing that the deflection angle $\theta$ (see Fig.~\ref{fig:setup_kinematics}a) is roughly uniform along the bundle, the projected acceleration can be approximated as $\ddot{w}_{B,\parallel} \simeq \ddot{w}_B \times \delta w_{B}/L$, under the assumption that $\theta \ll 1$.
As shown in Fig.~\ref{fig:setup_kinematics}e, the projected acceleration $\ddot{w}_{B,\parallel}$ is positive during most of the motion, indeed pushing the liquid towards the bundle tip.
Combining that kinematics with the fact that the longitudinal bundle permeability is much larger than the transversal one, we arrive at the conclusion that the flow of paint in the bundle mainly takes place in the direction parallel to the bristles and that the main driving force is the longitudinal projection $\ddot{w}_{B,\parallel}$ of the bundle acceleration.
%
\section{Flow regimes in the bundle} \label{sec:flow_regimes}
\begin{figure}[t]
    \centering
    \includegraphics[width = 0.95\linewidth]{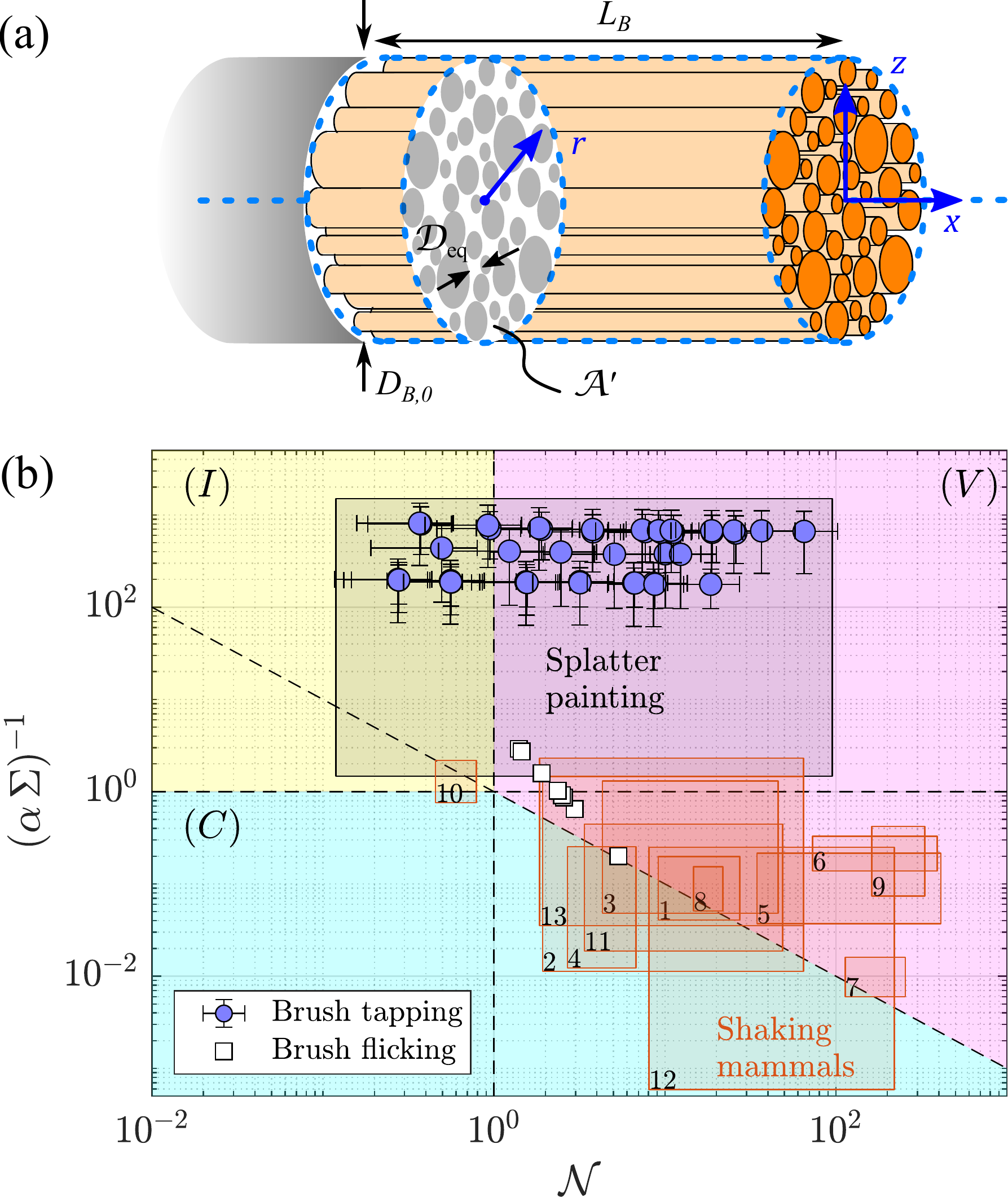} 
    \caption{(a) Sketch of the idealized fiber bundle considered here, along with the definitions of relevant geometrical parameters. --- (b) Regime map showing the dominant term in \eqref{eq:NSx_adim}, depending on the dimensionless parameters $(\alpha\Sigma)^{-1}$ and $\N$ defined in \eqref{eq:adim_nb}. Symbols show the location of our data for brush tapping and flicking. The gray shaded area shows the range of parameters where splatter painting artists are expected to work (see Section \ref{apdx:splatter_range} in SI). For comparison, orange rectangles mark the ranges of parameters for various shaking mammals, labeled by their number (see SI, Section \ref{apdx:mammals_range}).}
    \label{fig:bundle_flow_regimes}
\end{figure}

%
Since the driving force pushing paint out of a tapped brush has been identified, we now propose a  description of the liquid flow in the bundle.
As can be appreciated in Fig.~\ref{fig:setup_kinematics}b, the wet bundle keeps a roughly cylindrical shape during its oscillations.
We therefore model it as cylindrical slab of porous material of diameter $D_{B,0}$ and length $L_B$ (Fig.~\ref{fig:bundle_flow_regimes}a), in which all the pores (the spaces between fibers) are assumed to be parallel.
The cross-sectional area available for fluid flow, $\Aprime$, and the typical pore diameter, $\Dh$, are inferred from measurements of  $D_{B,0}$ (defined at the bundle foot), $L_B$, and distribution of bristle diameters in the brush (see SI, Section \ref{apdx:properties_porous}).
%
\subsection{Equation of motion for the fluid}
%
Let us now write an equation of motion for the incompressible and Newtonian liquid flowing in the idealized bundle depicted in Fig.~\ref{fig:bundle_flow_regimes}a. 
Momentum conservation applied in the (non-inertial) reference frame of the bundle reads
\begin{equation} \label{eq:NS}
    \rho \frac{\partial \Vec{v}}{\partial t} + \rho \Vec{v} \cdot \Vec{\nabla} \Vec{v} = - \Vec{\nabla}p + \mu \Vec{\nabla}^2 \Vec{v} - \rho \Vec{a}_0,
\end{equation}
where $\Vec{v}$ is the (three-dimensional) fluid velocity, $p$ the pressure and $\Vec{a}_0 = \ddot{w}_B (t) \, \Vec{e_Z}$ the linear (vertical) bundle acceleration in the reference frame of the laboratory (Fig.~\ref{fig:setup_kinematics}a).
The characteristic length scale in the transverse direction is $\Dh \sim 0.1~\milli\meter$, the hydraulic diameter of a pore in the bundle (see SI, Section \ref{apdx:properties_porous}). 
In the longitudinal direction, we take the characteristic length scale as the initial wet length of the bristles, $\ell_0$, that is to say, the initial liquid column length in the bundle. 
This magnitude is deduced from the liquid mass $m_0$ initially loaded into the brush bundle as $\ell_0 = m_0 / \rho \Aprime \sim 25~\milli\meter$.

Equation \ref{eq:NS} is projected onto the longitudinal direction of the bundle, which is expected to be the main flow direction. 
Under the approximation of slender flow $\varepsilon = \Dh /\ell_0 \ll 1$ and using the coordinate system $(r,x)$ tied to the bundle, we arrive at
\begin{equation} \label{eq:NSx}
\begin{split}
        \rho \frac{\partial v_x}{\partial t} + &\rho \left( v_x \frac{\partial v_x}{\partial x} + v_r \frac{\partial v_x}{\partial r} \right) = \\
        &-\frac{\partial p}{\partial x} + \frac{\mu}{r} \frac{\partial}{\partial r} \left( r \frac{\partial v_x}{\partial r} \right) - \rho \, \ddot{w}_{B,\parallel},
\end{split}
\end{equation}
where $v_x (r,x,t)$ and $v_r(r,x,t)$ denote the components of the fluid velocity in the longitudinal and radial directions, respectively, and $\ddot{w}_{B,\parallel} (x, t)$ is the longitudinal acceleration (Fig.~\ref{fig:setup_kinematics}e).

Equation \ref{eq:NSx} is non-dimensionalized using the oscillation half-period $t_c \sim 3~\milli\second$ as the characteristic timescale, and the capillary pressure $\sigma/\Dh$ as the pressure scale. 
The acceleration is scaled by a characteristic value $a_c = a_{\parallel}$ obtained from experimental measurements of $\ddot{w}_{B,\parallel}$ (see Section \ref{apdx:impact_params} of SI).
Setting the \textit{a priori} unknown longitudinal velocity scale $v_c$ to $v_c = \ell_0 / t_c$ to ensure dimensional consistency, we arrive at
\begin{equation} \label{eq:NSx_adim}
\begin{split}
        \alpha^{-1} &\left(\frac{\partial \bar{v}_{\bar{x}}}{\partial \bar{t}} + \bar{v}_{\bar{x}} \frac{\partial \bar{v}_{\bar{x}}}{\partial \bar{x}} + \bar{v}_{\bar{r}} \frac{\partial \bar{v}_{\bar{x}}}{\partial \bar{r}} \right) = \\
        &-\Sigma \frac{\partial \bar{p}}{\partial \bar{x}} + \N \alpha^{-1} \, \frac{1}{\bar{r}} \frac{\partial}{\partial \bar{r}} \left( \bar{r} \frac{\partial \bar{v}_{\bar{x}}}{\partial \bar{r}} \right) - \bar{a}_{\parallel},
\end{split}
\end{equation}
where dimensionless variables are denoted with a bar.
Three dimensionless control parameters arise:
\begin{equation} \label{eq:adim_nb}
     \alpha = \frac{a_c t_c^2}{\ell_0}; \quad  
     \N = \frac{\nu t_c}{\Dh^2}; \quad
     \Sigma = \frac{\sigma}{\rho a_c \ell_0 \Dh}.
\end{equation}
The number $\alpha$ is the dimensionless acceleration stemming from the choice of length scale $\ell_0$ and time scale $t_c$.
The number $\N$, which we call dimensionless viscosity, compares the viscous term to the inertial term in \eqref{eq:NSx_adim}.
For a periodic oscillating flow, this parameter would be equivalent to the Womersley number $Wo = \N^{-1/2}$.
The number $\Sigma$, which we call dimensionless surface tension, compares the capillary pressure gradient to the pressure gradient generated by the acceleration. 
For a gravity driven flow, $\Sigma$ would be the inverse of the Bond number. 
%
\subsection{Regime map}
%
Equation \ref{eq:NSx_adim} tells us that the momentum injected by the impact (last term on the right-hand side) is in part used to accelerate the liquid, in part dissipated by viscous friction, and in part used to deform liquid/gas interfaces.
Which of those mechanisms is dominant depends the relative size of the control parameters, evaluated through the ratio between the viscous and inertial terms, $\N$, and the ratio between the inertial and capillary terms, $(\alpha\Sigma)^{-1} = \rho \Dh \ell_0^2/\sigma t_c^2$.

Figure~\ref{fig:bundle_flow_regimes}b shows the parameter space $(\N, (\alpha\Sigma)^{-1})$, in which three zones can be distinguished:
\begin{itemize}
    \item For $\N < 1$ and $(\alpha\Sigma)^{-1}>1$ (yellow area), the inertial term in \eqref{eq:NSx_adim} dominates over the capillary and viscous terms, and the brush acceleration is mostly used to set the liquid in motion.
    \item For $\N > 1$ and $(\alpha\Sigma)^{-1}>1/\N$ (magenta area), the viscous term in \eqref{eq:NSx_adim} dominates over the capillary and inertial terms, and the brush acceleration is mostly used to overcome viscous friction.
    \item For $(\alpha\Sigma)^{-1} < \mathrm{min}(1, 1/\N)$ (cyan area), the capillary term in \eqref{eq:NSx_adim} dominates over the viscous and inertial terms, and the brush acceleration is mostly used to overcome the longitudinal capillary pressure gradient.
\end{itemize}
The range of parameters where we expect splatter artists to work is represented in Fig.~\ref{fig:bundle_flow_regimes}b by the gray shaded area, which was  calculated based on fluid properties of commercial and artist-made paints, and various brush geometries (see Sections \ref{apdx:splatter_artists} and \ref{apdx:splatter_range} in SI).
Our experiments - marked as symbols in the same map - fall into that area, mostly covering the viscous-dominated regime, and extending slightly towards the inertia-dominated domain. 
For comparison, we computed the parameter ranges (orange rectangles) corresponding to 13 species of shaking mammals, based on the shaking frequencies measured by Dickerson \textit{et al.} \cite{Dickerson2012}. 
In that case, viscous and longitudinal capillary forces are of similar order to magnitude, and both contribute to resisting water expulsion from hair clumps.
%
\section{How much liquid is ejected?}\label{sec:ejected_mass}
\subsection{Modeling the ejected mass} \label{ssec:modelling_mass}
%
\begin{figure}[t]
    \centering
    \includegraphics[width = \linewidth]{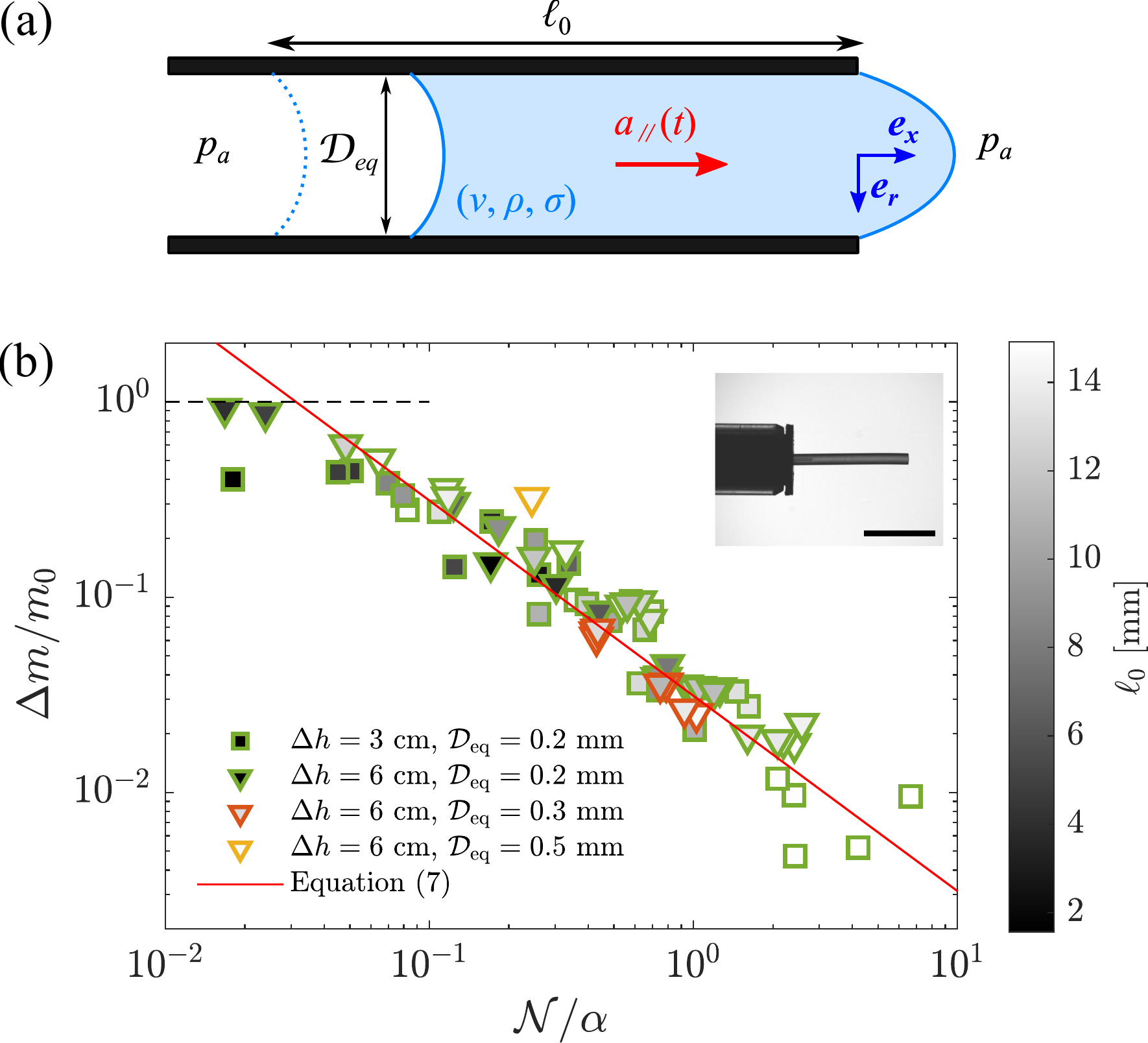}
    \caption{(a) Sketch of a further simplification of the bundle, consisting in a single cylindrical pore with diameter $\Dh$. --- (b) Liquid mass $\Delta m$ expelled from a single capillary tube, normalized by the initial mass $m_0$, as a function of the dimensionless parameter $\N /\alpha$ for various initial lengths of the liquid plug $\ell_0$ (grayscale), internal tube diameters $\Dh$ (symbol edge color) and drop heights $\Delta h$ (symbol shape). The red solid line represents the prediction in the viscous limit (~\eqref{eq:Dm_adim}), while the dashed line shows the inertial limit $\Delta m/m_0 = 1$. Scale bar on snapshot: $1~\centi\meter$.}
    \label{fig:single_pore_model}
\end{figure}
\begin{figure*}[t]
    \centering
    \includegraphics[width = \linewidth]{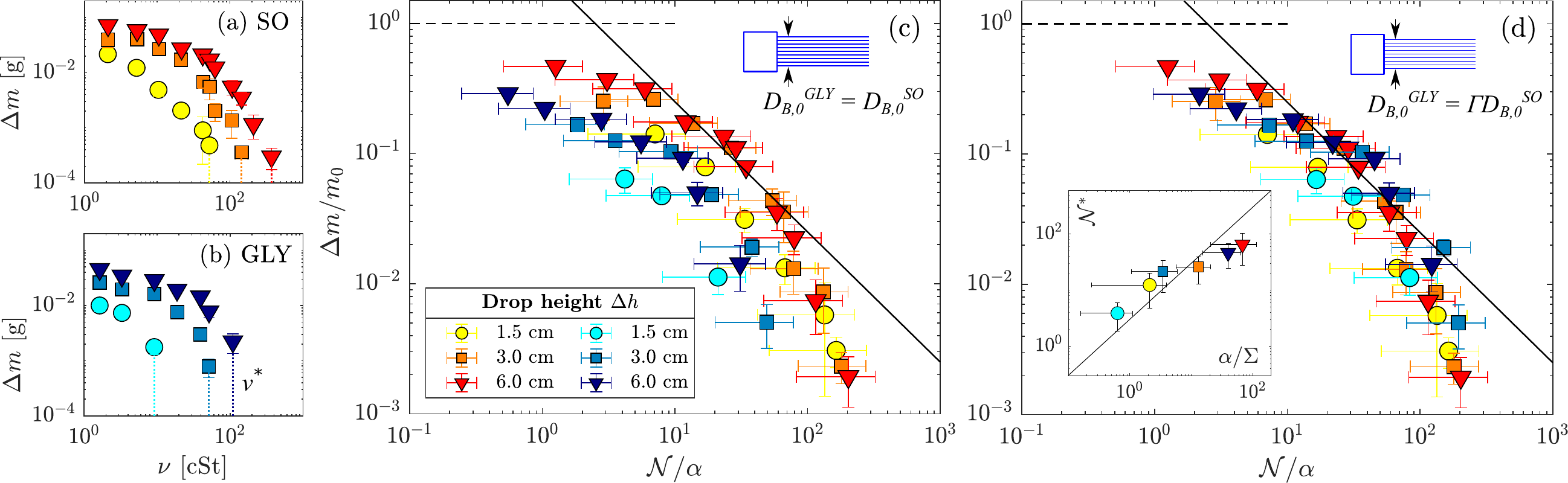}
    \caption{Mass $\Delta m$ ejected from a tapped brush as a function of the kinematic viscosity $\nu$ for (a) silicone oils and (b) water-glycerol mixtures and different drop heights. --- Relative mass loss $\Delta m / m_0$ as a function of the dimensionless parameter $\N /\alpha$ (symbols) for (c) identical wet bundle diameters for GLY and SO data, $D_{B,0}^{GLY} = D_{B,0}^{SO}$, and (d) a slightly tighter bundle for GLY, $D_{B,0}^{GLY} = \Gamma D_{B,0}^{SO}$ with $\Gamma = 0.93$. The solid line represents $\Delta m/m_0 = 2.5 \times (\N/\alpha)^{-1}$, while the dashed line shows the inertial limit $\Delta m/m_0 = 1$. --- Inset: Critical dimensionless viscosity $\N^{\ast} = \nu^{\ast} t_c/\Dh^2$, above which no measurable liquid detachment is observed, as a function of the dimensionless parameter $\alpha /\Sigma$. The solid line shows the scaling $\N^{\ast} \sim \alpha/\Sigma$ predicted by equation~\eqref{eq:condition_Nstar}, with a prefactor equal to $3$.}
    \label{fig:mass_plot}
\end{figure*}
The total mass of liquid $\Delta m$ leaving the brush bundle can be estimated by  solving \eqref{eq:NSx} for the velocity field in a complex geometry and with a time-dependent forcing term.
To make simplify our experimental observations, we first consider a model brush: a single cylindrical pore, modeled as an impermeable tube of diameter $\Dh$, as depicted in Fig.~\ref{fig:single_pore_model}a.
This setup amounts to assuming that there is no liquid flow between the bristles in the transverse direction.

Since most of our experimental data lie in the viscous-dominated regime (see Fig.~\ref{fig:bundle_flow_regimes}b), we first restrict our analysis to the viscous limit $\N \gg 1$ and $(\alpha\Sigma)^{-1} \gg 1$.
Under those assumptions, the velocity field in the cylindrical pore reduces to the Poiseuille flow
\begin{equation}
    v_x(r,x,t) = \frac{\ddot{w}_{B,\parallel}(x,t)}{4\nu}\left(\frac{\Dh^2}{4} - r^2 \right).
    \label{eq:Poiseuille}
\end{equation}
The mass of liquid expelled from the tip ($x = 0$) of the pore, introducing the time $t_f$ at which liquid ejection stops, can be expressed as
\begin{equation}
    \Delta m_{1} = \int_{0}^{t_f} \mathrm{d}t \, \int_{0}^{\Dh/2} \rho v_x(r, 0, t) \, 2\pi r \,  \mathrm{d}r \approx \frac{\pi}{128} \frac{\rho}{\nu}\Dh^4 a_c t_c,    
    \label{eq:Dm_1pore}
\end{equation}
where we approximated $\int_0^{t_f} \ddot{w}_{B,\parallel}(0,t) \, \mathrm{d}t \sim a_c t_c$. Therefore, we deduce that the total mass of liquid expelled from the brush is $\Delta m = (4\Aprime/\pi \Dh^2) \, \Delta m_{1}$.
That expression can finally be recast in dimensionless form as
\begin{equation}
    \frac{\Delta m}{m_0} \approx \frac{1}{32}\frac{\alpha}{\N},
    \label{eq:Dm_adim}
\end{equation}
where $m_0 = \rho \Aprime \ell_0$ is the initial liquid mass in the bundle. 
The expelled mass is expected to be proportional to the imposed acceleration, and inversely proportional to the liquid viscosity.

We now briefly turn our attention to the purely inertial limit ($\N \ll 1$ and $(\alpha \Sigma)^{-1} \gg 1$). 
Assuming a plug flow $v_x = v_x(t)$ and $v_r = 0$, the momentum conservation \eqref{eq:NSx} reveals that the velocity $v_x$ remains constant for times $t > t_f$. 
Liquid is then expected to flow out of the pores until they are completely empty, hence $ \Delta m / m_0 = 1$ in the purely inertial limit.

%
\subsection{Comparison to experimental data for a single pore} \label{ssec:exp_mass_single_pore}
%
Experiments with the model bundle, consisting of a single tube filled with water/glycerol solutions, provide a direct test of equation \eqref{eq:Dm_adim}. 
The dimensionless control parameter $\N /\alpha = \nu \ell_0/\Dh^2 t_c a_c$ is varied by changing the liquid viscosity $\nu$, the acceleration $a_c$ (corresponding to different drop heights $\Delta h$), the internal tube diameter $\Dh$, and the initial length of the liquid plug $\ell_0$.
Figure~\ref{fig:single_pore_model}b shows the expelled liquid mass $\Delta m$ ($=\Delta m_1$ for a single pore), normalized by the initial mass $m_0$, and plotted as a function of $\N /\alpha$.
The experimental data is in quantitative agreement with the prediction in the viscous limit, given by equation \eqref{eq:Dm_adim} (solid red line), except for the lowest values of $\N /\alpha$, where the expelled mass approaches the inertial limit $\Delta m/m_0 = 1$ (black dashed line).
%
\subsection{Comparison to experimental data for a brush} \label{ssec:exp_mass_brush}
%
Tapping experiments with real brushes are carried out for two liquid families, differing by their surface tension: silicone oils (Fig.~\ref{fig:mass_plot}a) and water/glycerol mixtures (Fig.~\ref{fig:mass_plot}b).
In both cases, we vary the liquid viscosity $\nu$ and the imposed acceleration $a_c$ (through the drop height $\Delta h$).
We estimate the pore diameter $\Dh$ and initial length of the liquid plug $\ell_0$, but cannot change them systematically due to the complex, uncontrolled geometry of commercial brush bundles.

The corresponding data is made dimensionless in Fig.~\ref{fig:mass_plot}c in the same fashion as for the case of a single pore (Fig.~\ref{fig:single_pore_model}b).
For a given liquid family, all experimental data collapse onto a single master curve, supporting our assumption that pores in the bundle do not interact at first order. 
Interestingly, the master curve for glycerol-water solutions (GLY) seems shifted as compared the one for silicone oils (SO). 
We attribute this shift to elasto-capillary effects \cite{Boudaoud2007, Py2007, Duprat2012, Ha2020}: surface tension deforms the bundle of flexible bristles differently for the two liquid families.
For the same dry brush, high-surface-tension GLY is expected to lead to a wet bundle diameter $D_{B,0}$ about $10\%$ smaller than the one for low-surface-tension SO (see Section \ref{apdx:properties_porous} in SI).
Figure~\ref{fig:mass_plot}d shows that a correction of that order of magnitude is indeed sufficient to collapse the GLY master curve onto the SO one.
Now comparing the dimensionless brush data to the prediction of our scaling analysis, we observe that experiments follow $\Delta m/m_0 \sim 2.5 (\N/\alpha)^{-1}$ (solid line in Figs.~\ref{fig:mass_plot}c,d) in an intermediate parameter range $\N/\alpha \sim 10^1-10^2$. 
This power law is the consistent with the prediction of equation~\eqref{eq:Dm_adim} (viscous limit), albeit with a larger prefactor.
For $\N/\alpha \lesssim 10^1$, the data seems to transition towards the inertial limit $\Delta m/m_0 = 1$ (dashed line in Figs.~\ref{fig:mass_plot}c,d)
On the contrary, for $\N/\alpha \gtrsim 10^2$, the expelled mass decreases faster than expected in the viscous limit, as we explain in the following.

A liquid drop of volume $\mathcal{V}$ hanging from a capillary of diameter $\Dh$ detaches if surface tension forces $F_{\mathrm{cap}} = \pi \sigma \Dh$ are no longer sufficient to balance body forces, in our case an acceleration force $F_{\mathrm{acc}} = \rho a_c \mathcal{V}$.
This happens when the drop volume exceeds a critical value $\mathcal{V}^{\ast} \sim \sigma \Dh/\rho a_c$, which is known as Tate's volume \cite{Tate1864} in the case where $a_c = g$ (the acceleration of gravity).
Following equation~\eqref{eq:Dm_1pore}, the time $t^{\ast}$ required to fill this critical volume $\mathcal{V}^{\ast}$ is given by
\begin{equation}
    \rho \mathcal{V}^{\ast} \sim \frac{\rho}{\nu}\Dh^4 a_c t^{\ast},  
\end{equation}
where we approximated $\int_0^{t^{\ast}} \ddot{w}_{B,\parallel}(0,t) \, \mathrm{d}t \sim a_c t^{\ast}$.
In splatter painting, the acceleration imposed to the liquid in the pores of the bundle is transient by nature, with a typical timescale $t_c$.
As a consequence, drop detachment can occur only if $t^{\ast} < t_c$, otherwise the acceleration goes back to zero before a volume $\mathcal{V}^{\ast}$ could be filled.
That condition for detachment can simply be recast as
\begin{equation}
   \nu < \frac{\rho a_c^2 t_c\Dh^3}{\sigma} =\nu^{\ast} \quad \text{\textit{i.e.}} \\
    \quad \N < \frac{\alpha}{\Sigma} = \N^{\ast}.
    \label{eq:condition_Nstar}
\end{equation}
For each data set in Fig.~\ref{fig:mass_plot}a,b, the rightmost point corresponds to the largest tested viscosity that yielded a measurable expelled mass. 
Using that point as a proxy for $\nu^{\ast}$, we are able to test the detachment condition by plotting $\N^{\ast} = \nu^{\ast} t_c/\Dh^2$ as a function of $\alpha/\Sigma$ for all the data sets (inset in Fig.~\ref{fig:mass_plot}d).
Experiments are found to be compatible with $\N^{\ast} = 4\alpha/\Sigma$ (solid line), in agreement with the prediction of~\eqref{eq:condition_Nstar}.
Note that $\nu^{\ast}$ strongly depends on the pore diameter, for which we used an average value $\Dh$. 
In the actual bundle, the pore size distribution is expected to translate into a distribution of values of $\nu^{\ast}$, explaining the gradual transition away from the trend $\Delta m / m_0 \sim (\N / \alpha)^{-1}$ as the detachment condition $\nu < \nu^{\ast}$ is met by fewer and fewer pores. Contrarily, in the experiments with a single pore (a tube), the flow is completely arrested when this condition is met.
%
\section{Conclusion} \label{sec:conclusion}
%
In summary, this work aimed at investigating the artistic technique of splatter painting from a fluid-mechanical perspective. 
High-speed imaging revealed the kinematics of the action (flicking or tapping) leading to paint ejection from the bristles of the brush.
Inspired by the empirical know-how developed by artists, we explored the influence of two key parameters – liquid viscosity and brush acceleration – on the quantity of liquid ejected from a brush subjected to tapping on its handle.
A dimensional analysis of the problem allowed us to identify that liquid inertia and viscosity may both contribute to balance the acceleration imparted to the liquid during splatter painting.
A simplified model in the viscous limit yielded a prediction for the mass of liquid ejected from the brush as a function of the control parameter $\N / \alpha$.
This prediction was found in qualitative agreement with experiments on tapped brushes, while additional experiments, performed with a simplified brush where the bundle is replaced by a single pore, showed quantitative agreement with the model.
Although the capillary pressure gradient was too weak to significantly resist acceleration in splatter painting, the data revealed two mechanisms through which surface tension had an effect on the amount of liquid expelled.
First, high surface tension liquids lead to tighter wet bundles, thereby reducing the space available for the liquid to flow between the bristles.
Second, a liquid drop detaches from the brush only if it has time to grow big enough to overcome surface tension forces during the acceleration phase. 
This leads to a surface tension dependent cut-off in viscosity, beyond which no mass is expelled at all.
We hope that this work will provide a useful framework for the study of other strongly accelerated flows in (anisotropic) porous media, found for instance in centrifugal drying \cite{anlauf2007, cheng2022}. Equally important, we hope it also inspires further studies of the complex and rich physical phenomena behind artistic painting techniques.
%
\section*{Materials and methods}\label{sec:exp}
%
\subsection*{Brush tapping experiments}
%
\paragraph{Materials}
%
In our tapping experiments, we use a set of identical commercial round brushes, purchased from a local crafts shop.
Each brush is composed of a wooden handle connected to a bundle of natural hairs (also called bristles) by a metallic ferrule.
The only modification we make to this brush is to trim the tip of the bundle, cutting all the bristles to a uniform length $L_B = 13.0 \pm 0.5~\milli\meter$ (see Fig.~\ref{fig:setup_kinematics}).
At the exit of the ferrule, the bundle has a diameter $D_{B,0} = 4.7 \pm 0.2~\milli\meter$.
Examination under the microscope reveals that the hairs are quite polydisperse, with an average diameter $\langle d \rangle = 169~\micro\meter \pm 53~\micro\meter$ (std), see Section \ref{apdx:properties_porous} in SI.
We assume that all of our brush copies have these same properties.
The handle length, measured from the clamping position to the tip of the ferrule, is set to $L_H = 112~\milli\meter$. 
The impact location of the striker, measured along the handle from the clamping position, is set to $x_i = 40 \pm 1~\milli\meter$.

We also perform a reduced set of tapping experiments with a simplified brush made of a single PTFE capillary tube -- emulating a pore in the bundle -- mounted at the tip of a rigid hollow acrylic tube ($10.1~\milli\meter$ in diameter) mimicking the handle.
Capillary tubes with the same outer diameter ($1.6~\milli\meter$) but various inner diameters ($0.2$, $0.3$ and $0.5~\milli\meter$) are used. 
The part of the tube sticking out of the handle (\textit{i.e.} the effective bundle length) is set to a fixed value of $L_B = 16~\milli\meter$.
For the simplified brush, the handle length and impact location of the striker are set to $L_H = 143~\milli\meter$ and $x_i = 47~\milli\meter$, respectively.

Most paints are non-Newtonian fluids, exhibiting shear-thinning, yield-stress or even viscoelastic properties \cite{Mysels1981, Lim2013}.
However, to be able to vary the fluid shear viscosity independently from the brush acceleration (and therefore shear rate), we use model liquids to perform the tapping experiments. 
Two families of Newtonian liquids are selected: silicone oils, denoted by prefix SO, and water-glycerol mixtures, denoted by prefix GLY.
Both these families allow us to vary viscosity over three orders of magnitudes while keeping density and surface tension roughly constant (see Section \ref{apdx:liq_properties} in SI).
%
\paragraph{Tapping setup}
%
We developed a custom-made experimental setup to mimic the tapping action produced by a splatter painting artist.
As shown in Fig.~\ref{fig:setup_kinematics}a, the handle is clamped in a horizontal position to simulate how the painter would hold the brush.
Controlled tapping is produced by dropping onto the handle a ring-shaped brass weight of mass $M = 85.65~\gram$, guided along a fixed vertical brass tube.
The impact generates a transverse wave that propagates in the handle all the way to the tip of the brush, where it is transmitted to the paint-loaded bristles (in the case of the real brush) or to the liquid-filled capillary tube (in the case of the simplified brush).
The resulting acceleration is adjusted by changing the height $\Delta h$ from which the weight is dropped (see Fig.~\ref{fig:setup_kinematics}a), from $\Delta h = 1.5$ to $6~\centi\meter$.
Note that the acceleration depends many other factors, such as striker mass $M$, handle length $L_H$, bundle length $L_B$ and location of impact $x_i$.
Those are not systematically varied in our study but their effect is discussed in Section \ref{apdx:impact_mechanics} of the SI.
%
\paragraph{Measurement setup and protocol}
%
For the real brush, the bundle is initially loaded manually with a fixed mass of liquid $m_0 = 155 \pm 5~\milli\gram$, which is kept constant throughout the experiments.
The mass $\Delta m$ of liquid ejected by the tapping action is measured by weighting the brush before and after impact with a high-precision scale (Sartorius, resolution $\pm 0.1~\milli\gram$).

For the simplified brush, the capillary tube is filled by dipping one end into a liquid pool and sucking a given amount with a syringe pump connected to the other end. 
The tube is then cut and mounted at the tip of the handle before performing the tapping.
The initial length of the liquid plug ranged from $\ell_0 = 1.6$ to $14.9~\milli\meter$.
The mass $\Delta m$ of liquid ejected by the impact is computed from the final length of the liquid plug $\ell_f$ measured in the transparent capillary tube, knowing the inner pore diameter and the liquid density.

During impact, the vertical displacements of the extremity of the handle and of the bundle, as well as the resulting liquid atomization, are visualized from the side and recorded by a high-speed camera (NAC Memrecam HX 3 or Photron Fastcam NOVA S12) at a frame rate of 30,000 fps.
A typical image sequence is presented in Fig.~\ref{fig:setup_kinematics}b.
Liquid detachment impedes the automatic detection of the bundle motion on the video footages (see snapshots in figure~3a of the main text).
This is the reason why we use here only the experimental videos for the most viscous liquids (SO-1000 and GLY-100), for which all the liquid remains trapped between the bristles.
We checked that the bundle kinematics is essentially independent of the viscosity of liquid loaded in the bristles, thus allowing us to extrapolate the results obtained on SO-1000 and GLY-100 to lower viscosities.
In-house image processing is subsequently applied to extract displacements and accelerations.

\subsection*{Brush flicking experiments}
%
One of the authors has replicated manually, in an approximate way, the flicking motion observed in the video of Octavio Moctezuma painting  the portrait displayed in Fig. \ref{fig:artwork_actions} and Movie S1. In the flicking experiments we have used a brush (bundle length approximately 20 mm) loaded with water, to guarantee that paint is easily ejected. Several flicking movements have been recorded, although here we only show a representative one for the sake of conciseness. The brush motion has been filmed with a high-speed camera Fastcam SA5 model 1300K-M3 at 6000 fps. The angles $\theta$ and $\varphi$ defined in Fig.~\ref{fig:kinematics_flicking}a have been measured using an in-house software written in Matlab. Then, the angular velocity $\omega$ and its time derivative $\dot{\omega}$ have been computed using central finite differences with a time step of 20 frames. 
%
%
\section*{Acknowledgments}
J.R.R. acknowledges financial support from Grant No. PID2023-146809OB-I00 funded by MICIU/AEI/10.13039/501100011033 and by ERDF/UE. J.R.R. and L.C. acknowledge funding from the Spanish MCIN/AEI/10.13039/501100011033 through Grant No. PID2020-114945RB-C21. 
This project has received funding from the European Union’s Horizon 2020 research and innovation programme under the Marie Sklodowska-Curie grant agreement No 882429 (L.C.). R.Z. acknowledges the financial of support of the Royce Family Professorship for teaching excellence at Brown University. 

We are very grateful to splatter painting artists Octavio Moctezuma and Caroline Champougny for sharing their expertise on this technique, as well as their paint recipes.
Many thanks as well to Luc\'{i}a Lacambra-Asensio for the measurements of splatter painting patterns.
%
\newpage
\onecolumn
\appendix
\begin{center}
%
\hrule \vspace{2mm}
{\LARGE \sf Supporting Information}
\vspace{2mm} \hrule
\vspace{5mm}
\end{center}
%
\section{Bundle geometry} \label{apdx:properties_porous}
%
In this section, we  experimentally characterize the geometry of the bundle of bristles at the tip of the brush used in tapping experiments.
We then deduce and discuss geometrical properties of the anisotropic porous medium used as an idealized representation of the bundle.
The values of relevant properties are listed in table~\ref{tab:bundle_props}.
%
\subsection{Dry bundle geometry}
%
The bundle length $L_B = 13.0 \pm 0.5~\milli\meter$ and base diameter (\textit{i.e.} at the end of the ferrule), $D_{B,0} = 4.70 \pm 0.05~\milli\meter$, are measured with a caliper.
In order to estimate the number of bristles and their diameter distribution, the bundle of one of the brush copies is severed at its base to be characterized.
The total number of hairs in the bundle is determined by weighting several known amount of hairs on a high precision scale (Sartorius, with a resolution of $\pm 0.1~\milli\gram$).
Measuring the total mass of all the bristles then allows us to deduce their number, which is found to be $N = 404 \pm 6$.

A reduced sample of hairs from the brush (about $25~\%$ of the total) is examined under the microscope to measure their diameters $d$. 
The experimental probability distribution $P(d)$ of bristles' diameters is then computed and characterized by its first and second moments, defined respectively as
\begin{align}
    \Mone &= \int_0^{\infty} d \, P(d) \, \mathrm{d}d \approx (177 \pm 3) \times 10^{-6} ~\meter, \label{eq:M1}\\
    \Mtwo &= \int_0^{\infty} d^2 P(d) \, \mathrm{d}d \, \approx (34 \pm 1)\times 10^{-9}~\meter^2. \label{eq:M2}
\end{align}
%
\subsection{Properties assigned to the idealized bundle}
%
The idealized bundle is assumed to have an overall cylindrical shape, with length $L_B$ and diameter $D_{B,0}$. 
It contains $N$ fibers whose diameter distribution is described by their first two moments, $\Mone$ and $\Mtwo$, given by equations~\eqref{eq:M1} and \eqref{eq:M2} respectively.
From those quantities, we can deduce the following properties for the idealized bundle.
\paragraph{Solid fraction} The solid volume fraction $\phi_s$ is defined as the total volume occupied by the fibers, normalized by the total volume of the bundle.
Denoting $\mathcal{A} = \pi D_{b,0}^2/4$ the total cross-sectional area of the bundle, we have
\begin{equation} \label{eq:solid_frac}
    \phi_s = \frac{\pi}{4} \frac{N \Mtwo}{\mathcal{A}} = \frac{N \Mtwo}{D_{b,0}^2}.
\end{equation}
The corresponding estimation of the bundle's solid fraction is provided in table~\ref{tab:bundle_props}.
%
\paragraph{Hydraulic diameter}
%
The equivalent (or hydraulic) diameter $\Dh$ of the bundle is defined as the diameter of a cylindrical pore that would have the same cross-sectional area $\mathcal{A}^\prime$ available for the liquid to flow and the same wet perimeter $\mathcal{P}$ as the whole bundle \cite{Gebart1992}.
Namely, 
\begin{equation} \label{eq:def_hydr_diam}
    \Dh = \frac{4 \mathcal{A}^\prime}{\mathcal{P}},
\end{equation}
where $\mathcal{A}^\prime = \mathcal{A} (1 - \phi_s)$ and $\mathcal{P} = \pi N \Mone$.
Using \eqref{eq:solid_frac} to replace $\mathcal{A}$, the hydraulic diameter can be recast as
\begin{equation} \label{eq:hydr_diam}
    \Dh = \frac{4 \mathcal{A} (1 - \phi_s)}{\mathcal{P}} = \frac{(1-\phi_s)}{\phi_s} \frac{\Mtwo}{\Mone}.
\end{equation}
The corresponding estimation of hydraulic diameter is provided in table~\ref{tab:bundle_props}.
Note that, in the case of a bundle made of identical fibers of radius $R$, the quantity $\Mtwo/\Mone$ reduces to $2R$, in accordance with  Gebart \cite{Gebart1992}.
For a monodisperse ordered bundle, the global solid fraction $\phi_s$,  \eqref{eq:solid_frac}, is also equal to the local solid fraction at the scale of one pore, which allows us to assume $\Dh$ as the hydraulic diameter of a single pore in the bundle. 
%
\paragraph{Permeabilities}
%
An anisotropic porous medium (such as that depicted in Fig. 3.a of the main text) has different permeabilities, depending if the flow is parallel or perpendicular to the fiber direction.
For a bundle of identical fibers of radius $R$, Gebart \cite{Gebart1992} derived analytical expressions for the longitudinal and transversal permeabilities, denoted $K_L$ and $K_T$, respectively:
\begin{align}
    K_L &= \frac{8R^2}{c} \frac{(1-\phi_s)^3}{\phi_s^2}, \label{eq:KL}\\
    K_T &= C_1 R^2 \left( \sqrt{\frac{\phi_{s,\mathrm{max}}}{\phi_s}} - 1\right)^{5/2}, \label{eq:KT}
\end{align}
where $c$, $C_1$ and $\phi_{s,\mathrm{max}}$ are constants that depend on the geometry of the fiber packing.
For a square packing, $c = 57$, $C_1 = 16/(9\pi\sqrt{2})$ and $\phi_{s,\mathrm{max}} = \pi/4$, while for a hexagonal packing $c = 53$, $C_1 = 16/(9\pi\sqrt{6})$ and $\phi_{s,\mathrm{max}} = \pi/(2\sqrt{3})$.
Interestingly, the ratio $\kappa$ between the longitudinal and transversal permeabilities is independent of the fiber radius $R$:
\begin{equation} \label{eq:perm_ratio}
    \kappa = \frac{K_L}{K_T} = \frac{8}{c C_1} \frac{(1-\phi_s)^3}{\phi_s^2 \left( \sqrt{\frac{\phi_{s,\mathrm{max}}}{\phi_s}} - 1\right)^{5/2}}.
\end{equation}
To the best of our knowledge, there are no expressions equivalent to equations \eqref{eq:KL}, and \eqref{eq:KT} in the literature for a bundle of polydisperse fibers.
However, since equation \eqref{eq:perm_ratio} does not depend on the fiber radius, we assume that the permeability ratio $\kappa$ for a polydisperse bundle is equal to the monodisperse one with the same solid fraction $\phi_s \pm \delta \phi_s$.
In that case, we estimate $\kappa \approx 5$ (respectively $\kappa \approx 10$) for fibers arranged in a hexagonal (resp. square) packing. Hence, we can conclude that the longitudinal flow dominates over the transversal one in the brush bundle, as assumed in our analysis.
%
\subsection{Wet bundle geometry}
\begin{figure*}[t]
    \centering
    \includegraphics[width = \linewidth]{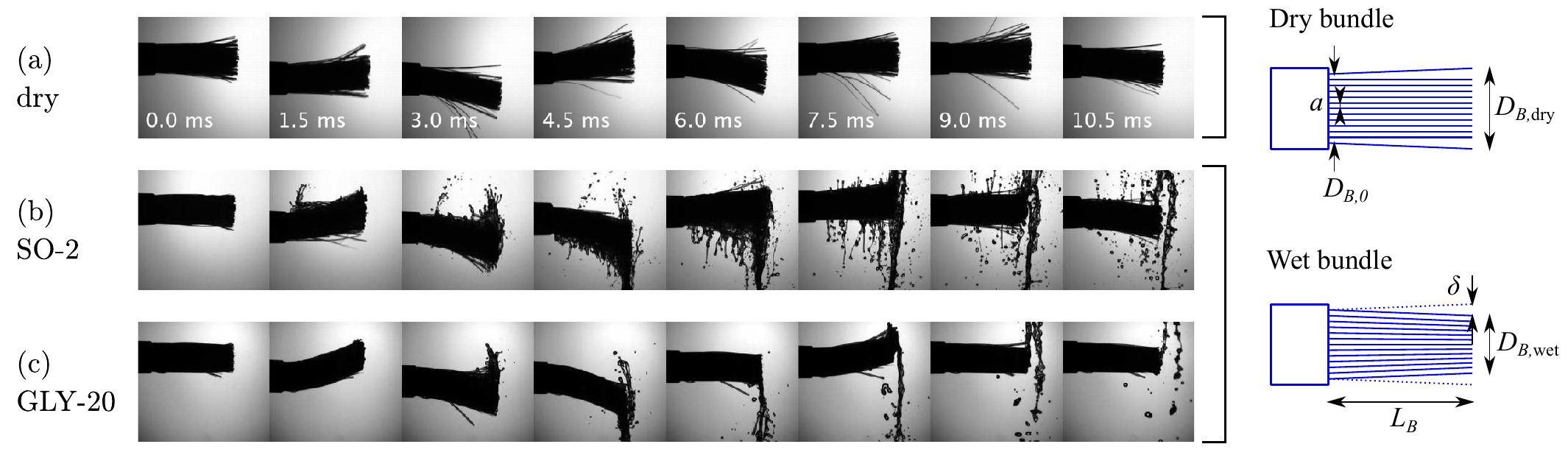} 
    \caption{Snapshots of bundle motion during an impact from a drop height $\Delta h = 3~\centi\meter$ (a) in the absence of liquid, (b) when loaded with SO-2 ($\nu = 2.1~\cSt$, $\sigma = 18.7~\milli\newton\per\meter$), (c) when loaded with GLY-20 ($\nu = 1.6~\cSt$, $\sigma = 71.7~\milli\newton\per\meter$). The timestamps are identical in the three sequences.}
    \label{fig:sigma_effect}
\end{figure*}

%
When a hairy structure is wet, capillary forces tend to bring the hairs closer together, an effect known as clumping \cite{Py2007, Duprat2012}.
This is the reason why the brush bundle at rest is tighter when it is wet (diameter at the tip $\Dwet$) as compared to when it is dry (diameter at the tip $\Ddry > \Dwet$), as can be appreciated in Figure~\ref{fig:sigma_effect}.

In a dynamic situation such as a bundle subjected to a sudden acceleration, cohesion is due to a combination of surface tension and viscous forces.
Like at rest, capillary bridges resist hair separation, as it would result in an increased interfacial area. 
Additionally, for two wet bristles to separate, the liquid between them must flow away, which becomes more and more difficult as viscosity increases. 

The snapshots in figure~\ref{fig:sigma_effect} show that, even for the smallest viscosities tested (SO-2 on panel b, GLY-20 on panel c), the wet hair bundle remains more cohesive during impact than in the dry case (panel a). 
Additionally, GLY-20, which has the largest surface tension ($\sigma = 71.7~\milli\newton\per\meter$) keeps the bundle tighter during motion as compared to SO-2 ($\sigma = 18.7~\milli\newton\per\meter$).
Those observations suggest that the wet bundle could be characterized by a diameter $\Dwet$ that is a function of surface tension.

\paragraph{Change of wet bundle diameter with surface tension}
%
In the following, we use scaling arguments to estimate how the wet bundle diameter is expected to change when using liquids with different surface tensions.
We denote $\delta$ the deformation of the wet hair bundle (as compared to the dry state), such that $\Dwet = \Ddry - 2\delta$, see Fig. \ref{fig:sigma_effect}.
That deformation results from a balance between the torque applied by capillary forces at the bundle tip, $F_{\mathrm{cap}} L_B \sim \sigma L_B^2$, and the elastic torque $F_{\mathrm{el}} L_B$ stemming from the resistance of the bundle to bending.
As shown in \cite{Ha2020}, this torque can be estimated as $F_{\mathrm{el}} L_B \sim EI \delta^2 / a L_B^2$
for a collection of bristles with Young modulus $E$, moment of inertia $I$ and separated by a distance $a$.
The diameter at the tip of the wet bundle is then obtained as
\begin{equation}
    \Dwet = \Ddry - 2\delta \sim \Ddry - 2 \sqrt{\frac{\sigma aL_B^4}{EI}}.
    \label{eq:Dwet_vs_sigma}
\end{equation}
Let us consider two identical brushes loaded with two different liquids, one with surface tension $\sigma_{\mathrm{SO}} = 20~\milli\newton\per\meter$ (representative of silicone oils) and the other with $\sigma_{\mathrm{GLY}} = 70~\milli\newton\per\meter$ (representative of glycerol/water solutions).
Using equation \eqref{eq:Dwet_vs_sigma}, the ratio $\Gamma$ between the diameters of the wet brushes can be expressed as
\begin{equation}
    \Gamma = \frac{\Dwet (\sigma_{\mathrm{SO}})}{\Dwet (\sigma_{\mathrm{GLY}})}%
    = \frac{1 - \epsilon}{1 - \beta \epsilon},
\end{equation}
where $\epsilon = 2\delta (\sigma_{\mathrm{SO}})/\Ddry$ and $\beta = \sqrt{\sigma_{\mathrm{GLY}}/\sigma_{\mathrm{SO}}} \approx 1.87$.
Taking advantage of the fact that $\epsilon \ll 1$ and of the definition $2\delta (\sigma_{\mathrm{SO}}) = \Ddry - \Dwet (\sigma_{\mathrm{SO}})$, we arrive at
\begin{equation}
    \Gamma%
    \approx 1 - (\beta - 1) \left( 1 -\frac{\Dwet (\sigma_{\mathrm{SO}})}{\Ddry}\right)%
    \approx 0.91,
    \label{eq:estimation_Gamma}
\end{equation}
where we estimated $\Dwet (\sigma_{\mathrm{SO}})/\Ddry \sim 0.9$, from the snapshots of a dry bundle and wet bundle (with silicone oil).
%
\paragraph{Effect of surface tension on bundle properties and data non-dimensionalization}
%
Table \ref{tab:bundle_props} shows that the bundle's geometrical properties are particularly sensitive to small variations in the bundle diameter ($<10~\%$), of the order of the one predicted by equation \eqref{eq:estimation_Gamma}.
This suggests that different hydraulic diameters $\Dh$ should be used to describe bundles loaded with silicone oils ($\sigma_{\mathrm{SO}} \approx 20~\milli\newton\per\meter$) and water/glycerol solutions ($\sigma_{\mathrm{GLY}} \approx 70~\milli\newton\per\meter$).

\begin{figure*}[t]
    \centering
    \includegraphics[width=1\linewidth]{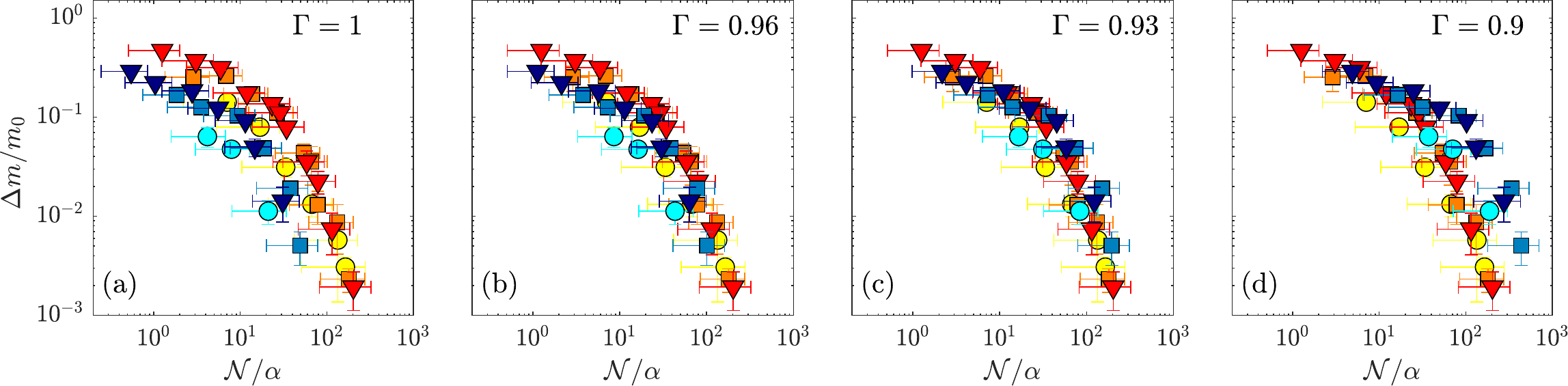}
    \caption{Non-dimensionalized mass $\Delta m / m_0$ as a function of the control parameter $\N / \alpha$ using slightly different bundle diameters $D_{b,0}$ and $\Gamma D_{b,0}$ for the silicone oils (orange shades) and water/glycerol solutions (blue shades), respectively.}
    \label{fig:dimensioness_data_varying_Gamma}
\end{figure*}

To test the consequences on the non-dimensionalization of our data, we set the bundle diameter to $D_{B,0}$ for liquids of the SO family and to $\Gamma D_{B,0}$ for liquids of the GLY family, using various values of $\Gamma$ in the range $0.9-1$. 
Figure~\ref{fig:dimensioness_data_varying_Gamma} presents corresponding non-dimensionalized data, $\Delta m /m_0$ as a function of the control parameter $\N / \alpha$. 
The color code is identical to the one in Fig. 5 of the main text (orange shades for SO, blue shades for GLY). 
We observe that a difference of only a few percents in bundle diameter is sufficient to shift the GLY data with respect to the SO data.
Good data collapse for the SO and GLY data is obtained with $\Gamma \approx 0.93$, which is consistent with a surface-tension-driven bundle tightening, as estimated by equation \eqref{eq:estimation_Gamma}.
\begin{table*}[t]
    \centering
    \caption{Geometrical properties assigned to the idealized bundle. The parameter $\Gamma$ accounts for variations of bundle diameter with respect to its nominal value $D_{B,0}$ (measured at the exit of the ferrule) under the action of capillary forces exerted by the liquid.}
        
    \begin{tabular}{llllll}
    \hline
    Property & $\Gamma = 1$ & $\Gamma = 0.96$ & $\Gamma = 0.93$ & $\Gamma = 0.9$ & Equation\\
    \hline
    Wet diameter $\Dwet$ & $4.70 \pm 0.05~\milli\meter$ & $4.51 \pm 0.05~\milli\meter$ & $4.37 \pm 0.05~\milli\meter$ & $4.23 \pm 0.05~\milli\meter$ & $D_{\mathrm{wet}} = \Gamma D_0$ \\
    Solid fraction $\phi_s$ & $0.63 \pm 0.04$ & $0.69 \pm 0.04$ & $0.73 \pm 0.04$ & $0.78 \pm 0.05$ & \eqref{eq:solid_frac}\\
    Equivalent diameter $\Dh$ & $113 \pm 18~\micro\meter$ & $89 \pm 14~\micro\meter$ & $71 \pm 12~\micro\meter$ & $54 \pm 9~\micro\meter$ & \eqref{eq:hydr_diam} \\
    \hline
    \end{tabular}
    \label{tab:bundle_props}
\end{table*}
%
\section{Kinematics of tapping} \label{apdx:impact_params}
%
Tapping experiments do not allow us to have a direct control over the acceleration of the brush.
In the tapping set-up, the initial energy input into the system is set by the drop height $\Delta h$ of the striker and its mass $M$.
How this translates into the accelerated motion of the bundle depends on several parameters, such as the location of impact, the geometry of the different parts of the brush (handle, bundle) and their mechanical properties.

In the main text, we use $\Delta h$ as the experimental control parameter modulating the acceleration, while keeping all other factors constant (the effect of some of them is discussed in section \ref{apdx:impact_mechanics}).
The bundle kinematics -- the vertical bundle acceleration $\ddot{w}_{B} (X,t)$, and its projection $\ddot{w}_{B,\parallel} (X,t)$ onto the longitudinal bundle direction -- is then measured as an output (see for instance Figure 2 in the main text).
In the following, we explain how we define and compute the observables characterizing the bundle kinematics.
Table~\ref{tab:ODM_acceleration} summarizes their values for the two different copies of the same brush, used for tapping experiments with the (SO) and (GLY) families respectively.

\paragraph{Characteristic timescale} 
The characteristic timescale $t_c$ of the flow is set by the oscillations of the bundle during the impact. 
Since both upwards and downwards motion may cause paint detachment, we define $t_c$ as the average half-period of oscillation of the bundle deflection $\delta w_B (t)$, as shown in Fig.~2.c in the main text.
%
\paragraph{Characteristic accelerations} 
The typical longitudinal acceleration $a_{\parallel}$ driving the flow along the bundle is defined as the time integral of $\ddot{w}_{B,\parallel} (X,t)$ during the first full period of oscillation:
\begin{equation}
    a_{\parallel} = \frac{1}{2 t_c} \int_0^{2 t_c} \ddot{w}_{B,\parallel} (X_{\mathrm{tip}},t) \, \mathrm{d}t,
\end{equation}
and evaluated at the bundle tip $X=X_{\mathrm{tip}}$, where most of the liquid expulsion occurs.
Similarly, the typical \emph{vertical} acceleration $a_0$ of the bundle can be defined as
\begin{equation}
    a_0 = \frac{1}{2 t_c} \int_0^{2 t_c} |\ddot{w}_{B} (X_{\mathrm{tip}},t)| \, \mathrm{d}t.
\end{equation}
Note that, due to limitations of the bundle tracking algorithm, we set $X_{\mathrm{tip}} \lesssim L_B$ for the real brush, while $X_{\mathrm{tip}} = L_B$ for the single-pore simplified brush.
\begin{table*}[t]
    \centering
    \caption{
    Observable  parameters characterizing bundle kinematics for two different copies of the same brush (one for the SO family and one for the GLY family), at location $X_{\mathrm{tip}} = 0.85 L_B$.
    }
    
    \begin{tabular}{lcccc}
    \hline
    Liquid family & Drop height & Oscillation time & Longitudinal acceleration & Vertical acceleration \\
     & $\Delta h$ ($\centi\meter$) & $t_c$ ($\milli\second$) & $a_{\parallel}$ ($\times 10^3~\meter\per\second^2$) & $a_0$ ($\times 10^3~\meter\per\second^2$) \\
    \hline
    \multirow{3}{*}{Brush (SO)} & $1.5$ & $3.0 \pm 0.9$ & $0.21 \pm 0.02$ & $1.71 \pm 0.09$ \\
                          & $3$ & $2.3 \pm 0.4$ & $0.70 \pm 0.01$ & $3.26 \pm 0.02$ \\
                          & $6$ & $2.2 \pm 0.5$ & $1.64 \pm 0.04$ & $5.01 \pm 0.07$ \\
    \hline
    \multirow{3}{*}{Brush (GLY)} & $1.5$ & $2.2 \pm 0.4$ & $0.34 \pm 0.03$ & $2.50 \pm 0.08$ \\
                                 & $3$ & $2.1 \pm 0.5$ & $0.77 \pm 0.01$ & $3.71\pm 0.07$\\
                                 & $6$ & $2.2 \pm 0.3$ & $2.6 \pm 0.2$ & $6.1 \pm 0.2$\\
    \hline
    \end{tabular}
 
    \label{tab:ODM_acceleration}
\end{table*}
%
\section{Impact mechanics} \label{apdx:impact_mechanics}
\begin{figure}[t]
\centering
\includegraphics[width=10cm]{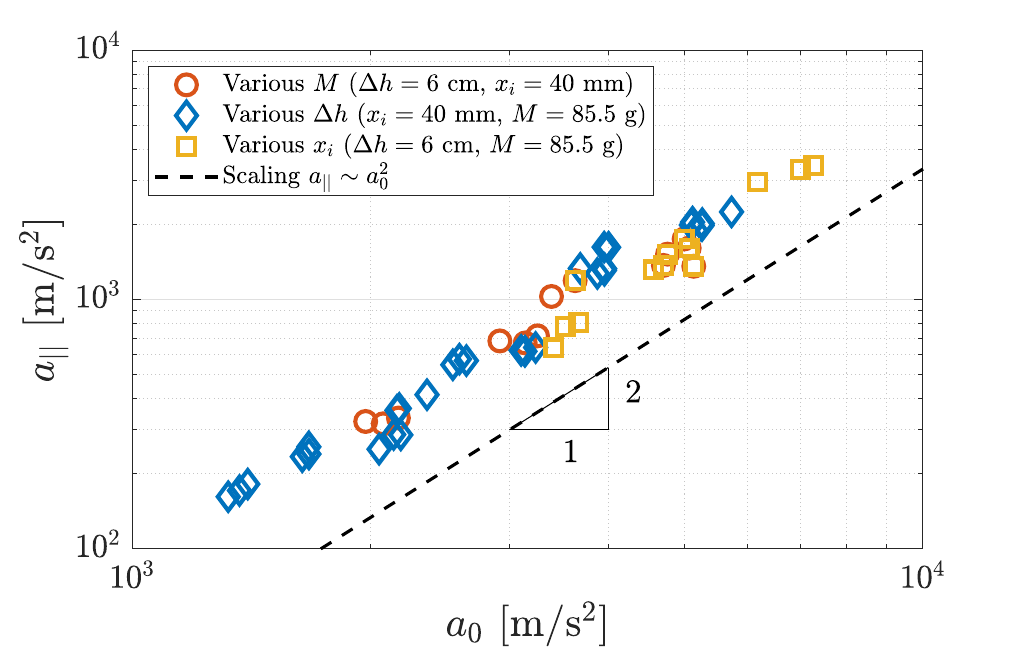}
\caption{\label{fig:ac_vs_a0}Characteristic longitudinal acceleration $a_{\parallel}$ as a function of the vertical one, $a_0$, for different experiments obtained varying the drop height $\Delta h$, striker mass $M$, and impact location $x_i$, respectively, while keeping the rest of the parameters constant. The solid line is a power law, $a_{\parallel} \sim a_0^2$, meant as a visual aid. Accelerations were computed at $X_{\mathrm{tip}} = 0.75 L_B$.}
\end{figure}

%
In the main text, the parameter we use to vary the acceleration imparted to the brush is the height $\Delta h$ from which the striker is dropped. 
However, many other factors may affect this acceleration, such as brush properties or impact conditions.
In this section, we develop and validate a simple solid mechanical model allowing us to relate these parameters to the produced acceleration.
%
\paragraph{Relation between vertical and longitudinal acceleration}
%
As explained in the main text, for a tapping action, the flow of paint inside the fiber bundle is mainly driven by the projection of the bundle's vertical acceleration, $a_0$, along its longitudinal axis, $a_{\parallel}$. 
Assuming that the bundle behaves as a linear beam, its deflection with respect to the horizontal direction, $\theta$, must be proportional to $a_0$ (see Fig. 2a in main text). 
Thus, we expect the longitudinal projection $a_{\parallel}$ to scale as
\begin{equation}
    a_{\parallel} \sim a_0\,\sin\theta \sim a_0^2.
\end{equation}
This argument is supported by the measurements shown in Fig. \ref{fig:ac_vs_a0} where $a_{\parallel}$ is plotted as a function of $a_0$ for different sets of experiments obtained impacting a same type of brush with strikers of different mass, $M$, dropped from different heights $\Delta h$, and at different impact locations $x_i$.\\
Fig. \ref{fig:ac_vs_a0} shows that, for a given brush, the relationship between $a_{\parallel}$ and $a_0$ is independent of the impact parameters ($\Delta h$, $M$, $x_i$), as all the points collapse on a single band of data. 
This allows us to directly map the dependency of $a_{\parallel}$ with the different control parameters onto the one of $a_0$, which we model in the next paragraph.
%
\paragraph{Modeling the vertical acceleration $a_0$}
%
Below, we use classical beam theory to model the motion of the impacted brush, following the approach proposed by Goldsmith \cite{goldsmith1998impact}.
The brush is assumed to be a single uniform cylindrical beam of diameter $D_H$ (handle diameter), length $ L = L_H + L_B$ and linear mass density $\lambda$.
The vertical acceleration of the tip of the brush, $a_0$, is computed as a function of the problem parameters, assuming small deformations.

\begin{figure}[t]
\centering
\includegraphics[width=\linewidth]{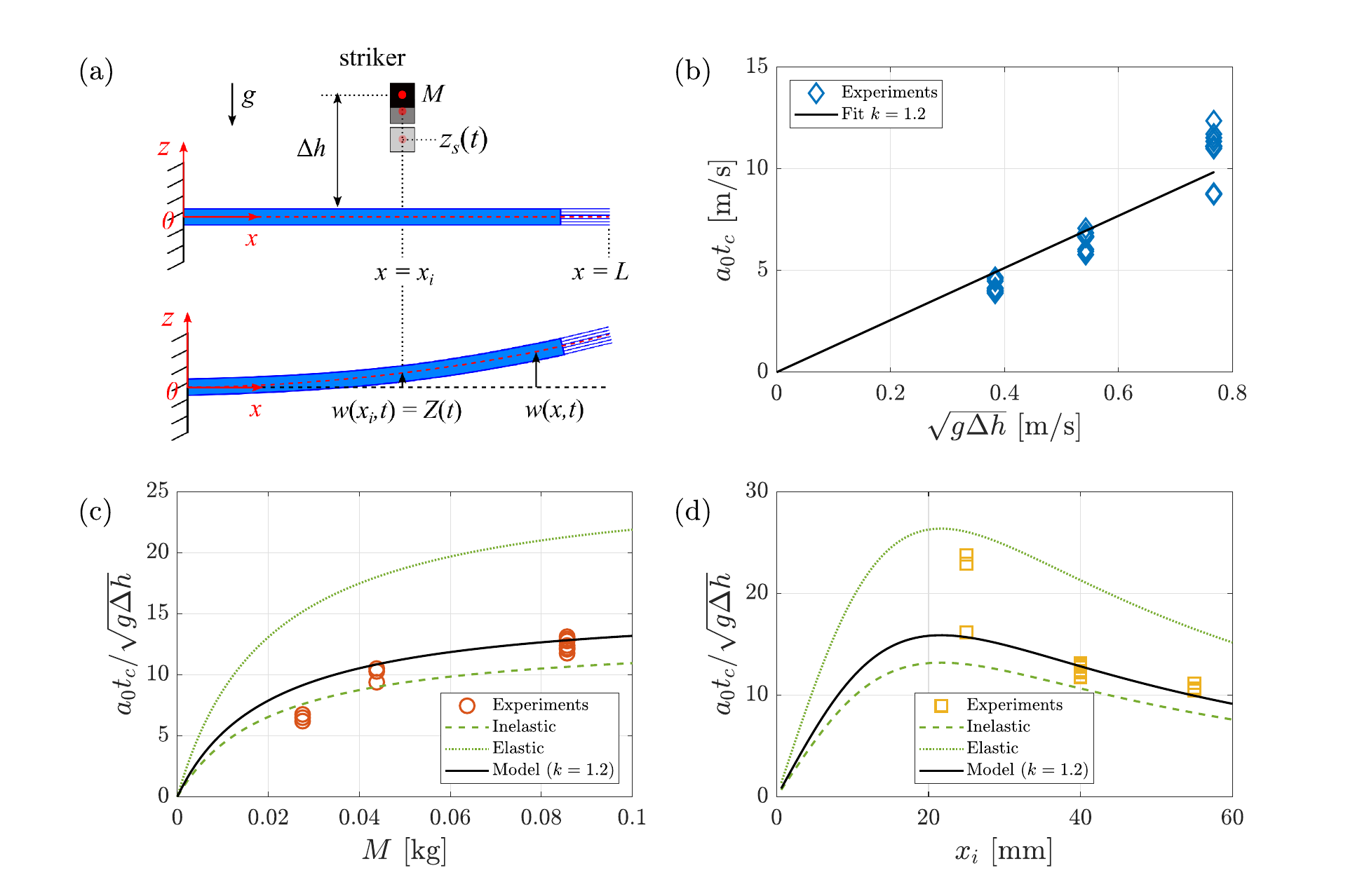}
\caption{(a) Sketch with notations used in the modeling of the handle motion upon impact of a striker of mass $M$. --- (b) Vertical acceleration of the brush's tip $a_0$ times the brush oscillation time $t_c$ as a function of $\sqrt{g\Delta h}$ for $x_i = 4~\centi\meter$ and $M = 85.66~\gram$. The solid line is a fit of the data with equation~\eqref{eq:a0_from_Deltah} using $k$ as the only fitting parameter.  --- Normalized vertical acceleration of the handle's tip as a function of (c) the striker's mass $M$ (for fixed $x_i = 4~\centi\meter$ and $\Delta h = 6~\centi\meter$), and (d) the location of impact $x_i$ (for fixed $M = 85.5~\gram$ and $\Delta h = 6~\centi\meter$). The solid lines are the predictions of equation~\eqref{eq:a0_from_Deltah} with the value of $k = 1.2$ fitted in panel b with a different data set. The dotted and dashed lines show the elastic and inelastic limits, respectively. Accelerations were computed at $X_{\mathrm{tip}} = 0.75 L_B$.
}
\label{fig:model_SolidMech}
\end{figure}

In this model, a point striker of mass $M$ hits the beam perpendicularly at a distance $x_i$ from the clamping, as depicted in Fig.~\ref{fig:model_SolidMech}a. The key assumption is that the beam's vertical displacement is given by
\begin{equation}
    w(x, t) = Z(t)\,W(x),
    \label{eq:deflection_w}
\end{equation}
where $Z(t)$ is the beam's vertical position at the impact location, $x = x_i$ and $W(x)$ is the static beam deformation in response to a constant point force applied at $x=x_i$, but scaled in such a way that $W(x_i) = 1$. 
For a uniform cantilever beam of length $L$ clamped at the origin:
\begin{equation}
    W(x) = \frac{-x^3/6 + x_i x^2/2 + {\cal H}[x-x_i](x-x_i)^3/6}{x_i^3/3},
\end{equation}
where ${\cal H}[x]$ is the Heaviside function.

\noindent The beam velocity at $x_i$ upon the impact, $\dot{Z}_{0}$, can be obtained for two limiting cases: inelastic and elastic collision.
In both cases, we must look for a generalized momentum that is preserved across the impact.
This can be readily done using Lagrangian mechanics.
In terms of the two coordinates defining the impact dynamics, namely the vertical coordinate of the striker $z_s$ and the deflection of the beam at the impact point, $Z$, the kinetic energy of the system reads
\begin{equation}
    E_K = \frac{1}{2}M\dot{z}_s^2 + \frac{1}{2} \dot{Z}^2 \lambda \int_0^{L}W(x)^2\,\mathrm{d}x.
    \label{eq:kinetic_energy}
\end{equation}
Here, we recall that $\lambda$ is the linear density of the brush, assumed constant.
The Lagrange equations of the system, integrated over the short duration of the impact \cite{goldsmith1998impact}, result in:
\begin{equation}
    \Delta \left(\frac{\partial E_K}{\partial \dot{z}_s}\right) = F_s, \; \mathrm{and} \; \Delta \left(\frac{\partial E_K}{\partial \dot{Z}}\right) = -F_s,
\end{equation}
where the operator $\Delta$ stands for variation of the magnitude during the impact and $F_s$ is the contact percussion force between the striker and the beam. Notice that, since the parametrization we have chosen (Eq. (\ref{eq:deflection_w})) already imposes $w(0) = \partial w(0) / \partial x = 0$, we do not need to introduce any percussion force/torque at the clamping to impose the boundary conditions there. Adding the two Lagrange equations together we get,
\begin{equation}
    \Delta\left(\frac{\partial E_K}{\partial \dot{z}_s} + \frac{\partial E_K}{\partial \dot{Z}}\right) = 0.
\end{equation}
Thus, the quantity in the parenthesis is the generalized momentum that is preserved at the impact. 
Evaluating this expression for the kinetic energy defined in Eq. (\ref{eq:kinetic_energy}), we finally get:
\begin{equation}
    M \dot{z}_{s,0^{-}} = M \dot{z}_{s,0^{+}} + \dot{Z}_{0} \lambda \int_0^{L} W(x)^2\,\mathrm{d}x,
\end{equation}
where $\dot{z}_{s,0^{-}}$ is the speed of the striker before the impact, and $\dot{z}_{s,0^{+}}$ and $\dot{Z}_{0}$ are the speed of the striker and the beam at the impact point right after, respectively.

In the inelastic impact, besides the conservation of the generalized momentum we also impose the equality of the velocity of the striker and beam after the impact, $\dot{z}_{s,0^{+}}=\dot{Z}_{0}$, as $z_s(t) = Z(t)$ for $t>0$. Conversely, in the elastic case we must complement the conservation of the generalized momentum with the conservation of kinetic energy:
\begin{equation}
    \frac{1}{2} M \dot{z}^2_{s,0^{-}} = \frac{1}{2} M \dot{z}^2_{s,0^{+}} + \frac{1}{2} \dot{Z}^2_{0} \lambda \int_0^{L} W(x)^2\,\mathrm{d}x.
\end{equation}
Finally, after some algebra we obtain:
\begin{equation}
    \frac{\dot{Z}_{0}}{\dot{z}_{s,0^{-}}} = \frac{k}{1 + (\lambda/M) \int_0^{L} W(x)^2\, \mathrm{d}x}.
    \label{eq:speed_after_impact_inelastic}
\end{equation}
where $k = 1$ for the inelastic case and $k = 2$ for the elastic one.

Now we need to relate the vertical acceleration of the beam's tip to the initial velocity at the impact point, $\dot{Z}_{0}$. 
If the motion of the beam was purely sinusoidal, with a half-period $t_c$, then
\begin{equation}
    a_0 = \left|\frac{\mathrm{d}\dot{Z}}{\mathrm{d}t}(t=t_c/2)\right|W(L) = \dot{Z}_{0}\frac{\pi}{t_c}W(L) = \frac{k \sqrt{g\Delta h} \, }{1 + (\lambda/M) \int_0^{L} W(x)^2\, \mathrm{d}x}\frac{\pi}{t_c}W(L), 
    \label{eq:a0_from_Deltah}
\end{equation}
assuming that the striker falls freely, thus $\dot{z}_{s,0^{-}} = \sqrt{g \Delta h}$. 
In practice, the constant $k$ can be used as a free parameter, as the impact will neither be perfectly elastic nor inelastic.
%
\paragraph{Comparison to experimental data}
%
To find the value of the constant $k$, we use the results of an experimental data set obtained with a striker of mass $M = 85.66~\gram$, impacting at a fixed location $x_i = 40~\milli\meter$, but dropping from different heights $\Delta h$. 
From the mean beam diameter $D_H = 6.5$ mm, and assuming a typical volumetric density for soft wood $\rho_{\mathrm{wood}} = 700~\kilo\gram\per\meter^3$, we estimate the value of the linear density $\lambda = 23~\gram\per\meter$. 
In figure \ref{fig:model_SolidMech}b we show the vertical accelerations $a_0$ measured at the tip of the bundle for various drop heights $\Delta h$. 
The solid line is a linear fit of the experimental data, allowing us to deduce the constant $k = 1.2$ from equation~\eqref{eq:a0_from_Deltah}.

\noindent Once $k$ is known, equation~\eqref{eq:a0_from_Deltah} gives a prediction for the dependency of $a_0$ on other problem parameters, such as the striker mass $M$ and impact location $x_i$. 
This prediction is plotted as a solid line in figure \ref{fig:model_SolidMech}c, and compared to independent experimental data with varying $M$ ($x_i$ and $\Delta h$ being fixed).
The dotted and dashed lines show the elastic and the inelastic limiting cases, respectively.
We find that the dependency of $a_0$ with the striker mass, $M$, is predicted reasonably well by the model without further fitting (\textit{i.e.} simply using the value of $k$ obtained from figure~\ref{fig:model_SolidMech}b). 
Similarly, figure \ref{fig:model_SolidMech}d, we plot the acceleration $a_0$ measured in response to impacts of a given striker falling from given height at different locations $x_i$ along the brush. The model also correctly captures the effect of the impact location $x_i$. 

The above calculations and experimental results allow us to reach two important conclusions. First, for a same type of brush, the acceleration that drives the flow inside the bundle, $a_{||}$, can be simply related to the vertical acceleration $a_0$ at the tip of the brush, regardless the impact conditions.
Second, the dependency of the acceleration $a_0$ on the parameters of the impact can be rationalized using simple and classical tools from beam theory considering reasonable assumptions: uniform properties of the beam, perfectly sinusoidal oscillation of the brush after the impact, speed of the striker before the impact equal to its maximum free-fall speed, and modeling the beam deflection using expression \eqref{eq:deflection_w}. 
The agreement is nearly quantitative, as only one free constant is needed.
Once this constant is found experimentally, the model successfully predict the results of other independent experiments. Finally, it it worth noting that the elastic and inelastic curves contain no free parameter, and yet they still describe the bundle's tip acceleration almost quantitatively.

\section{Liquid preparation and properties} \label{apdx:liq_properties}
%
\subsection{Model liquids}
%
Tapping experiments are carried out at ambient temperature, with average $\langle T \rangle = 21.6~\degreecelsius$.
Table \ref{tab:liq_properties} summarizes the physical properties of the liquids at that temperature.
%
Silicone oils SO-2, 5, 10, 20, 50, 100, 200, 350, 1000 were used off the shelf. 
In addition, silicone oils SO-40, 70, 150 were obtained by blending commercially available oils using the Grunberg-Nissan mixing rule \cite{Grunberg1949}.
The kinematic viscosities of the blends were subsequently measured with an Ostwald viscometer.
Aqueous glycerol solutions are prepared mixing pure glycerol ($\geq 99~\%$) and ultrapure (Milli-Q) water in various proportions.
Solutions are labelled GLY-$\xi$, where $\xi$ stands for the mass fraction of glycerol in $\%$, while GLY-100 simply denotes pure glycerol.
The density and viscosity of these solutions were computed based on $\xi$ and the ambient temperature using the correlation-based calculator developed by Andreas Volk and Chris Westbrook \cite{GLYcalculator, Volk2018}.

\begin{table}[t]
    \centering
    \caption{Properties at $\langle T \rangle = 21.6~\degreecelsius$ of the model liquids used in tapping experiments: silicone oils (prefix SO) and water-glycerol solutions (GLY).}
    
    \begin{tabular}{lccc}
    \hline
    Liquid code & Kinematic viscosity $\nu$ & Density $\rho$ & Surface tension $\sigma$\\
     & $(\cSt)$ & $(\kilo\gram\per\cubic\meter)$ & $(\milli\newton\per\meter)$\\
    \hline
    SO-2 & $2.1$ & $876$ & $18.7$\\
    SO-5 & $5.3$ & $921$ & $19.7$\\
    SO-10 & $11$ & $938$ & $20.1$\\
    SO-20 & $21$ & $953$ & $20.6$\\
    SO-40 & $43$ & $958$ & $20.7$\\
    SO-50 & $53$ & $963$ & $20.8$\\
    SO-70 & $63$ & $967$ & $20.9$\\
    SO-100 & $107$ & $969$ & $20.9$\\
    SO-150 & $147$ & $970$ & $21.0$\\
    SO-200 & $213$ & $971$ & $21.0$\\
    SO-350 & $373$ & $973$ & $21.1$\\
    SO-1000 & $1070$ & $974$ & $21.2$\\
    \hline
    GLY-20 & $1.6$ & $1045$ & $71.7$\\
    GLY-40 & $3.2$ & $1098$ & $70.0$\\
    GLY-60 & $8.8$ & $1152$ & $68.5$\\
    GLY-70 & $17$ & $1178$ & $67.0$\\
    GLY-78 & $35$ & $1200$ & $67.4$\\
    GLY-80 & $48$ & $1208$ & $67.1$\\
    GLY-87 & $100$ & $1225$ & $66.7$\\
    GLY-100 & $970$ & $1260$ & $63.4$\\
    \hline
    \end{tabular}

    \label{tab:liq_properties}
\end{table}
%
\subsection{Paints}
%
In addition to the model liquids described in table~\ref{tab:liq_properties}, used for controlled tapping experiments, we also determined the physical properties of various paints used by splatter painting artists.
These include dilutions of acrylic paint, concentrated watercolors, and traditional tempera (a mixture of egg yolk with linseed oil and water). 
The measured values of kinematic viscosity $\nu$, surface tension $\sigma$ and density $\rho$ are reported in table~\ref{tab:paint_properties}.

We asked artist Caroline Champougny to prepare acrylic paint solutions suitable for splatter painting.
The solutions were made using commercial acrylic paint (FW, Daler Rowney) of various colors diluted in ultrapure (Milli-Q) water.
%
We also characterized a tempera, prepared under the guidance of artist Octavio Moctezuma using egg yolk (1 part), linseed oil (1 part) and dionized water (4 parts). 
This paint was used for the painting shown in Fig. 1 of the main article and its corresponding video. 
Tempera paint exhibits a shear thinning and weakly viscoelastic behavior, with viscosities between $0.07$ and $0.014~\pascal \cdot \second$ for shear rates ranging from $0.1$ to $500~\second^{-1}$; the elastic relaxation time is around  $1~\milli\second$. 
%
Finally, we determine the properties of some commercial paints used off-the-shelf by splatter painting artists, see Table \ref{tab:splatter_artists}: liquid watercolors (Blick), high flow acrylic paint (Golden), and waterproof India ink (Blick). 
\begin{sidewaystable}
    \centering
    \caption{Properties of some paints used for splatter painting, either prepared following the artist's instructions (acrylic paint solutions and tempera) or used directly from the bottle.}
    \begin{tabular}{llcccc}
    \hline
    Paint type & Color & Concentration & Kinematic viscosity $\nu$ & Density $\rho$ & Surface tension $\sigma$\\
    & & & $(\cSt)$ & $(\kilo\gram\per\cubic\meter)$ & $(\milli\newton\per\meter)$\\
    \hline
    Acrylic paint (FW, Daler Rowney) & Zinc White (104) & $19~\%$ & $5.6$ & $1010$ & $57$\\
    diluted in water     & Azo Yellow Medium (269) & $7.9~\%$ & $1.9$ & $1001$ & $75$\\
         & Ultramarine (504) & $8.3~\%$ & $3.8$ & $1004$ & $78$\\
         & Ultramarine violet (507) & $15~\%$ & $5.4$ & $1014$ & $57$\\
         & Oxide Black (735) & $7.7~\%$ & $2.3$ & $1003$ & $66$\\
    \hline
    Tempera & - & - & $7.0-14.1$ & $993.3$ & $59.6$\\
    \hline
    Liquid watercolor (Blick) & Yellow & - & $54.9$ & $1000.6$ & $59.8$\\
    High flow acrylic paint (Golden) & Black & - & $1.9-186$  & $1072$ & $47.9$\\
    India ink (Blick) & Black & - & $2.0$ & $1035$ & $39.6$\\
    \hline
    \end{tabular}
    \label{tab:paint_properties}
\end{sidewaystable}
%
\section{Examples of artists who use the splatter painting technique} \label{apdx:splatter_artists}
%
Beyond the two artists who collaborated with this work -- Octavio Moctezuma and Caroline Champougny -- and several famous artists such as Jackson Pollock, 
Sam Francis, Robert Motherwell, Adolph Gottlieb, to name a few, splatter painting turns out to be a rather common artistic practice.
This is evidenced by the numerous artists and art studios who have documented their use of the technique on digital platforms, such as YouTube. 
The non-exhaustive list in Table \ref{tab:splatter_artists} shows some examples, along with the conditions of splatter painting reported in the video (accessible by clicking on the source).
%
\begin{sidewaystable}
  \centering
  \caption{Compilation of some artists and art studios who currently use the splatter painting technique. In all cases, brushes of different shapes and sizes are used. All links were last accessed on July 19th, 2026.}
  \label{tab:splatter_artists}
  \small
 \begin{tabular}{llll}    
     \hline
    \textbf{Source} & \textbf{Substrate} & \textbf{Action} & \textbf{Paint and preparation} \\
    \midrule
    \href{https://www.youtube.com/watch?v=dapBBbNPQhY&t=128s&ab_channel=LuckyScootersOfficial}{Lucky Scooters} &  metal  & flinging & Valspar Enamel Spray Paint, not prepared \\
    \href{https://artfulparent.com/splatter-painting-with-kids-crazy-fun-for-all-ages/}{The artful parent} &  paper, canvas, cardboard &  flinging & Liquid watercolor (Color Splash!), not prepared \\
    \href{https://www.youtube.com/watch?v=Yhnk3Emik9M&ab_channel=RachelFroudArt}{Rachel Froud Art} &  paper & bristle flicking, tapping & acrylics, water diluted \\
    \href{https://www.youtube.com/watch?v=doom3lYSMIk&ab_channel=WildlifeinWatercolour}{Wildlife in Watercolor} &  paper & bristle flicking, tapping & Windsor \& Newton Cotman Watercolour, water diluted  \\
    \href{https://www.youtube.com/watch?v=2_kt2vXjCkQ&ab_channel=JayLeePainting}{Jay Lee Painting} &  paper  & flinging, tapping & Schminke watercolor, water diluted  \\
    \href{https://www.youtube.com/watch?v=xjYrKkQJLro&ab_channel=DoodlesandScribbles}{Doodles and Scribbles} &  paper  & flinging, bristle flicking & acrylics, water diluted  \\
    \href{https://www.youtube.com/watch?v=xjYrKkQJLro&ab_channel=DoodlesandScribbles}{Doodles and Scribbles} & paper & flinging & Washable broadline markers (Liqui-Mark), not modified \\
    \href{https://www.youtube.com/watch?v=nsQxG5jX9vQ&ab_channel=KarenRiceArt}{Karen Rice Art} &  paper & tapping, bristle flicking & Sennelier Watercolor, water diluted \\
    \href{https://www.youtube.com/watch?v=t8wf54QXuL4&ab_channel=PaintAlongWithSkye}{Paint Along With Skye} &  canvas  & tapping & acrylic paint, water diluted \\
    \href{https://www.youtube.com/watch?v=-xMxgcSRTqA&ab_channel=TheMindofWatercolor}{The Mind of Watercolor} &  paper & bristle flicking, tapping & watercolor paint,  water diluted \\
    \href{https://www.youtube.com/watch?v=gveXwIjQo_w&ab_channel=Ms.Covart}{Ms. Covart} &  paper & bristle flicking, tapping & watercolor paint,  water diluted \\
    \href{https://www.youtube.com/watch?v=nQUmNielwrw&ab_channel=theartsherpa}{The ArtSherpa} &  canvas & tapping & high-flow ink, fluid acrylics (GOLDEN), not prepared \\
    \href{https://www.youtube.com/watch?v=129suvTDvyc&ab_channel=WatercolorMisfit}{WatercolorMisfit} &  paper  & tapping & Dr. Ph. Martin's concentrated water color,  water diluted \\
    \href{https://www.youtube.com/watch?v=GKSm8pKzYo4&ab_channel=FineArt-Tips}{Fine Art-Tips} &  paper & tapping & Watercolor paint Schmincke,  water diluted \\
    \href{https://www.youtube.com/watch?v=68zFcuG8U5c&ab_channel=MarsupialPudding}{MarsupialPudding} &  canvas &  tapping & acrylic paint, water diluted \\
    \href{https://www.youtube.com/watch?v=60IAsUSvsPs&ab_channel=StayCreativePaintingwithRyanO%27Rourke}{Stay Creative Painting with Ryan O'Rourke} &  canvas  & bristle flicking & acrylic paint,  water diluted\\
    \href{https://www.youtube.com/watch?v=27op9ik-OCI&ab_channel=EmmaJaneLefebvre}{Emma Jane Lefebvre} &  paper & tapping & Dr. Ph. Martin's Bleedproof White Ink,  water diluted \\
    \href{https://www.youtube.com/watch?v=TNH3YQQe-Ag&ab_channel=thefrugalcrafterLindsayWeirich}{thefrugalcrafter Lindsay Weirich} &  paper & flinging & watercolor paint,  water diluted \\
    \href{https://www.youtube.com/watch?v=EbvbmFz-Q6s&ab_channel=CommandoDesigns}{Commando Designs} &  plastic objects & flinging & Krylon Colormaster Spray Paint, not prepared \\
    \bottomrule
  \end{tabular}
\end{sidewaystable}
%
\section{Parameter range for splatter painting artists} \label{apdx:splatter_range}
\begin{figure}[t]
    \centering
    \includegraphics[width=\linewidth]{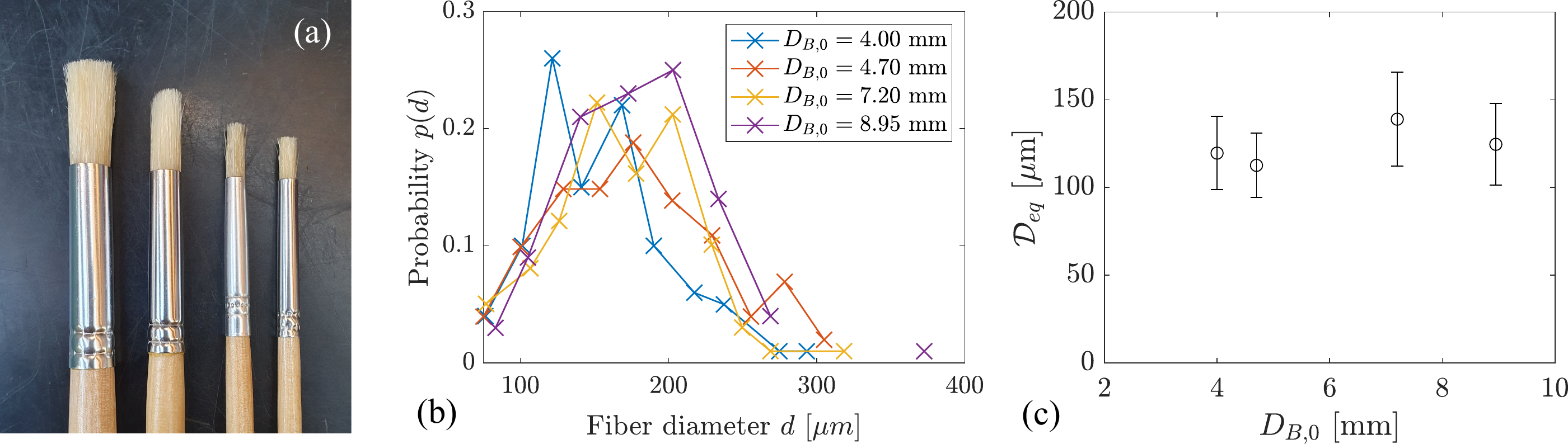}
    \caption{\textbf{Bundle properties of brushes of the same family but different sizes.} (a) Brushes used in these measurements. The diameter of the thickest ferrule is 9 mm (b) Probability size distribution of the bristle diameter $d$ for the four brushes shown in (a). (c) Equivalent hydraulic diameter of the pores between bristles corresponding to the probability size distribution shown in (b).}
    \label{fig:Dh_vs_Db0}
\end{figure}

%
In this section, we explain how we estimate the range of dimensionless parameters $\N = \nu t_c/\Dh^2$ and $(\alpha\Sigma)^{-1} = \rho \Dh \ell_0^2/\sigma t_c^2$ where splatter painting artists work.
Two kinds of parameters need to be determined: on the one hand paint properties (kinematic viscosity $\nu$, density $\rho$ and surface tension $\sigma$), and on the other hand parameters depending either directly or indirectly on brush geometry (equivalent diameter $\Dh$, length of liquid column $\ell_0$ and, as we develop below, characteristic oscillation timescale $t_c$).
In this section, we estimate $\N = 0.1-95$ and $(\alpha\Sigma)^{-1} = 1.5-1500$, which is the range marked as a shaded area in the regime map in Fig. 3.b of the main text.
\paragraph{Paint properties}
The physical properties of different paints used by splatter painting artists are reported in table~\ref{tab:paint_properties}. 
Based on those values, we set $\nu = 1-100~\cSt$, $\sigma = 45-75~\milli\newton\per\meter$ and $\rho = 1000-1070~\kilo\gram\per\meter^3$ as the typical ranges for paint kinematic viscosity, surface tension and density, respectively.
\paragraph{Equivalent pore diameter}
%
We characterized round brushes with natural hairs (MILAN series 512) following the procedure described in Section~\ref{apdx:properties_porous}. 
The bristle diameter and size distribution were found to be independent of brush size, and so was the typical equivalent diameter $\Dh \sim 100~\micro\meter$ (\ref{fig:Dh_vs_Db0}).
\paragraph{Liquid column length}
%
The length $\ell_0$ of the liquid column in the bundle depends on the amount of paint loaded by the artist, which is in general unknown. 
However, the length of the bundle can be used as a reasonable order of magnitude: $\ell_0 \sim L_B \sim 10-30~\milli\meter$ for this brush series.
\paragraph{Oscillation timescale}
%
We now evaluate how the brush oscillation timescale $t_c$ varies when changing the brush geometry.
To do so, we start from the observation that, in the tapping set-up, the oscillation of the bundle is essentially driven by the one of the handle.
We use the elastic beam theory \cite{goldsmith1998impact} to estimate the frequency $\omega_c$ of the first oscillation mode of the handle, modeled as a uniform cantilever beam:
\begin{equation}
    \omega_c^2 = \frac{EI}{\lambda}\left(\frac{c\pi}{L_H}\right)^4,
\end{equation}
where $c=0.596864$ is a dimensionless constant, $\lambda$ is the linear density of the beam, $E$ the Young modulus, $I=\pi D_H^4/64$ the moment of inertia of the cross-section (assumed to be circular of diameter $D_H$), and $L_H$ the beam's length, measured from its clamping position.
The half-period of oscillation $t_c = \pi/\omega_c$ is then simply expressed as
\begin{equation}
    t_c = \frac{4}{\pi c^2} \sqrt{\frac{\rho_{\mathrm{wood}}}{E}} \frac{L_H^2}{D_H},
    \label{eq:tc_beam_theory}
\end{equation}
where $\rho_{\mathrm{wood}} \approx 700~\kilo\gram\per\meter^3$ is the density of the wooden handle.
The only remaining unknown is the value of the Young modulus.
Using the value of $t_c \approx 2.3~\milli\second$ measured for our brush of dimensions $L_H = 112~\milli\meter$ and $D_H = 6.5~\milli\meter$, we estimate $E \approx 6300~\mega\pascal$, which in reasonable agreement with the tabulated values for a soft wood \cite{ross2010}.
Knowing $E$, Equation~\eqref{eq:tc_beam_theory} then allows us to estimate the oscillation timescale for a variety of brush geometries.
Using typical ranges $L_H \sim 100-200~\milli\meter$ and $D_H~\sim 5-10~\milli\meter$, for the handle length and diameter, respectively, we arrive at $t_c \sim 1-10~\milli\second$.
%
\section{Parameter range for shaking mammals} \label{apdx:mammals_range}
%
Similarly to splatter painting, mammals shaking to dry their fur also produce an acceleration-driven flow in an anisotropic porous medium (hair clump).
To examine the force balance at play, we estimate in this section the range of dimensionless parameters $\N = \nu t_c/\Dh^2$ and $(\alpha\Sigma)^{-1} = \rho \Dh \ell_0^2/\sigma t_c^2$ corresponding to shaking mammals.

The kinematic viscosity $\nu$, density $\rho$ and surface tension $\sigma$ are those of water.
Consistently with our definition in the case of splatter painting, the characteristic timescale is taken as the oscillation half-period: $t_c = 1/(2f)$, where $f$ is the shaking frequency measured by Dickerson \textit{et al.} for various mammal species \cite{Dickerson2012}. 
The liquid length column $\ell_0$ may be approximated by the typical hair length for a given species. 
As illustrated by equation~\eqref{eq:hydr_diam}, the hydraulic diameter $\Dh$ depends on the hair diameter distribution and on the solid fraction in the hair clump, both of which are unknown in the case of shaking mammals. 
For a given species, we therefore simply use the hair diameter as a proxy for the equivalent diameter $\Dh$ in the hair clump. 

Table~\ref{tab:shaking_mammals} contains a list of 13 mammal species for which we were able to find the shaking frequency, typical hair length and hair diameter in the literature. 
For each species, the data in table~\ref{tab:shaking_mammals} is used to estimate the ranges of $\N$ and $(\alpha\Sigma)^{-1}$ shown in Fig.~3.b of the main text.

\begin{sidewaystable}
    \centering
    \caption{Shaking frequencies, hair length and hair diameter of for various shaking mammals. Shaking frequencies are all obtained from \cite{Dickerson2012}. Whenever specified in the corresponding reference (rightmost column), the ranges of hair length and diameter include the two kinds of hairs commonly found in mammals: guard hairs and underfur.
}
    \begin{tabular}{lllcccc}
    \hline
    Number & Common name & Scientific name & Shaking frequency & Hair length & Hair diameter & Reference\\
    & & & ($\hertz$) & ($\milli\meter$) & ($\micro\meter$) & \\
    \hline
        1 & mouse & \textit{Mus musculus} & $27-31$ & $6.2-9.4$ & $26-42$ & \cite{Dry1926} \\
        2 & rat & \textit{Rattus norvegicus} & $16-20$ & $6-24$ & $22-114$ & \cite{Fish2002}\\
        3 & domestic cat & \textit{Felis catus} & $9.4$ & $17-49$ & $34-111$ & \cite{Lehmann2020}\\
        4 & river otter & \textit{Amblonyx cinereus} & $10.2$ & $5-18$ & $85-135$ & \cite{Kuhn2010} \\
        5 & Siberian husky & \textit{Canis lupus familiaris} & $5.4-5.8$ & $39-48$ & $15-50$ & \cite{Lapinski2014}\\
        6 & chow & \textit{Canis lupus familiaris} & $5.0$ & $79-80$ & $16-37$ & \cite{Lapinski2014} \\
        7 & kangaroo & \textit{Macropus rufus} & $4.9$ & $15-20$ & $20-30$ & \cite{Dawson2017}\\
        8 & Labrador retriever & \textit{Canis lupus familiaris} & $4.3-4.6$ & $26-39$ & $73-86$ & \cite{Vaishnav2021}\\
        9 & Boer goat & \textit{Capra hircus} & $7.7$ & $40-80$ & $14-20$ & \cite{Zhang2022} \\
        10 & Kunekune pig & \textit{Sus scrofa} & $8.2$ & $27-40$ & $278-366$ & \cite{Phadmacanty2023}\\
        11 & black bear & \textit{Ursus americanus} & $4.1$ & $20-50$ & $50-190$ & \cite{Brown1942}\\
        12 & Sumatran tiger & \textit{Panthera tigris sumatrae} & $4.3$ & $5-45$ & $23-120$ & \cite{Kitpipit2013}\\
        13 & brown bear & \textit{Ursus arctos horribilis} & $4$ & $30-100$ & $44-260$ & \cite{Brown1942}\\

    \hline
    \end{tabular}

    \label{tab:shaking_mammals}
\end{sidewaystable}
%
\newpage
\bibliography{biblio}
\bibliographystyle{unsrt}
        %

\end{document}